%% file: ms.tex
\documentclass{emulateapj}

\shorttitle{Spitzer observations of the complex S254-S258}
\shortauthors{Chavarr\'{\i}a et al.}

\newcommand{\be}{\begin{equation}}
\newcommand{\ee}{\end{equation}}
\newcommand{\bd}{\begin{displaymath}}
\newcommand{\ed}{\end{displaymath}}
\newcommand{\bi}{\begin{itemize}}
\newcommand{\ei}{\end{itemize}}
\newcommand{\bfig}{\begin{figure}}
\newcommand{\efig}{\end{figure}}
\newcommand{\bc}{\begin{center}}
\newcommand{\ec}{\end{center}}
\newcommand{\hii}{{H\scriptsize{II}}}
\newcommand{\ai}{$\alpha_{\textrm{\tiny{IRAC}}}$}
\newcommand{\complex}{S254-S258}

\begin{document}


\title{Spitzer observations of the Massive star forming complex \complex: \\structure and evolution}


\author{L. Chavarr\'{\i}a\altaffilmark{1,2}, L. Allen\altaffilmark{1}, J. L. Hora\altaffilmark{1}, C. Brunt\altaffilmark{3}, and G. G. Fazio\altaffilmark{1}}





\altaffiltext{1}{Harvard-Smithsonian Center for Astrophysics, 60 Garden Street, Cambridge, MA02138, USA}
\altaffiltext{2}{Departamento de Astronom\'{\i}a, Universidad de Chile, Camino del Observatorio 1515, Las Condes, Santiago, Chile}
\altaffiltext{3}{The School of Physics, University of Exeter, The Queens Drive, Exeter, Devon, UK EX4 4QL}


\begin{abstract}
We present Spitzer-IRAC, NOAO 2.1meter-Flamingos, Keck-NIRC, and FCRAO-SEQUOIA observations of the massive star forming complex \complex, covering an area of $25\times 20$ arc-minutes. Using a combination of the IRAC and NIR data, we identify and classify the young stellar objects (YSO) in the complex. We detect 510 sources with near or mid IR-excess, and we classify 87 Class I, and 165 Class II sources. The YSO are found in clusters surrounded by isolated YSO in a low-density distributed population. The ratio of clustered to total YSO is 0.8. We identify six new clusters in the complex. One of them, G192.63-00, is located around the ionizing star of the \hii~region S255. We hypothesize that the ionizing star of S255 was formed in this cluster. We also detect a southern component of the cluster in \hii~region S256. The cluster G192.54-0.15, located inside \hii~region S254 has a V$_{\mathrm{LSR}}$ of 17~km~s$^{-1}$ with respect to the main cloud, and we conclude that it is located in the background of the complex. The structure of the molecular cloud is examined using $^{12}\mathrm{CO}$ and $^{13}\mathrm{CO}$, as well as a near-IR extinction map. The main body of the molecular cloud has V$_{\mathrm{LSR}}$ between 5 and 9~km~s$^{-1}$. The arc-shaped structure of the molecular cloud, following the border of the \hii~regions, and the high column density in the border of the \hii~regions support the idea that the material has been swept up by the expansion of the \hii~regions.
\end{abstract}





\section{Introduction}
The Spitzer space telescope has proved to be a useful tool to study the spatial distributions and evolutionary states of young stars in regions of star formation. The combination of Spitzer-IRAC and near-IR (NIR) data allow us to identify the young stellar objects (YSO) in those regions. Several studies involving IRAC+NIR data have appeared recently, most of them on low-mass regions and in part due to the Legacy and GTO surveys \citep{pad08,cha07,har07b,por07,jor06,gut07,all07}. In this work, we present Spitzer-IRAC, 2.1meter-Flamingos, and Keck-NIRC imaging covering the whole complex of high mass star formation S254-S258. We use the IRAC+NIR data to identify and classify the YSO in the complex. We also combine our data with $^{12}\mathrm{CO}$ and $^{13}\mathrm{CO}$ emission in the region, in order to understand how massive stars interact with the interstellar medium. In this paper, we develop the method of analysis and present the results for the \complex~complex. In a following paper, we will repeat our analysis in other massive star forming regions and compare the results with the studies for low mass star forming regions (Chavarr\'{\i}a et al. 2008, in preparation).

Following is a brief description of the \complex~complex. In \S~2 we describe our data. Results of the analysis, including the identification of YSO, an analysis of their spatial distributions, and an investigation of the structure of the molecular cloud, are present in \S~3. In \S~4 we discuss individual clusters, the ages of the \hii~regions, and the overall picture. Our conclusions are presented in \S~5.

\subsection{Sharpless 254-258}
\begin{figure}
\epsscale{0.9}
\plotone{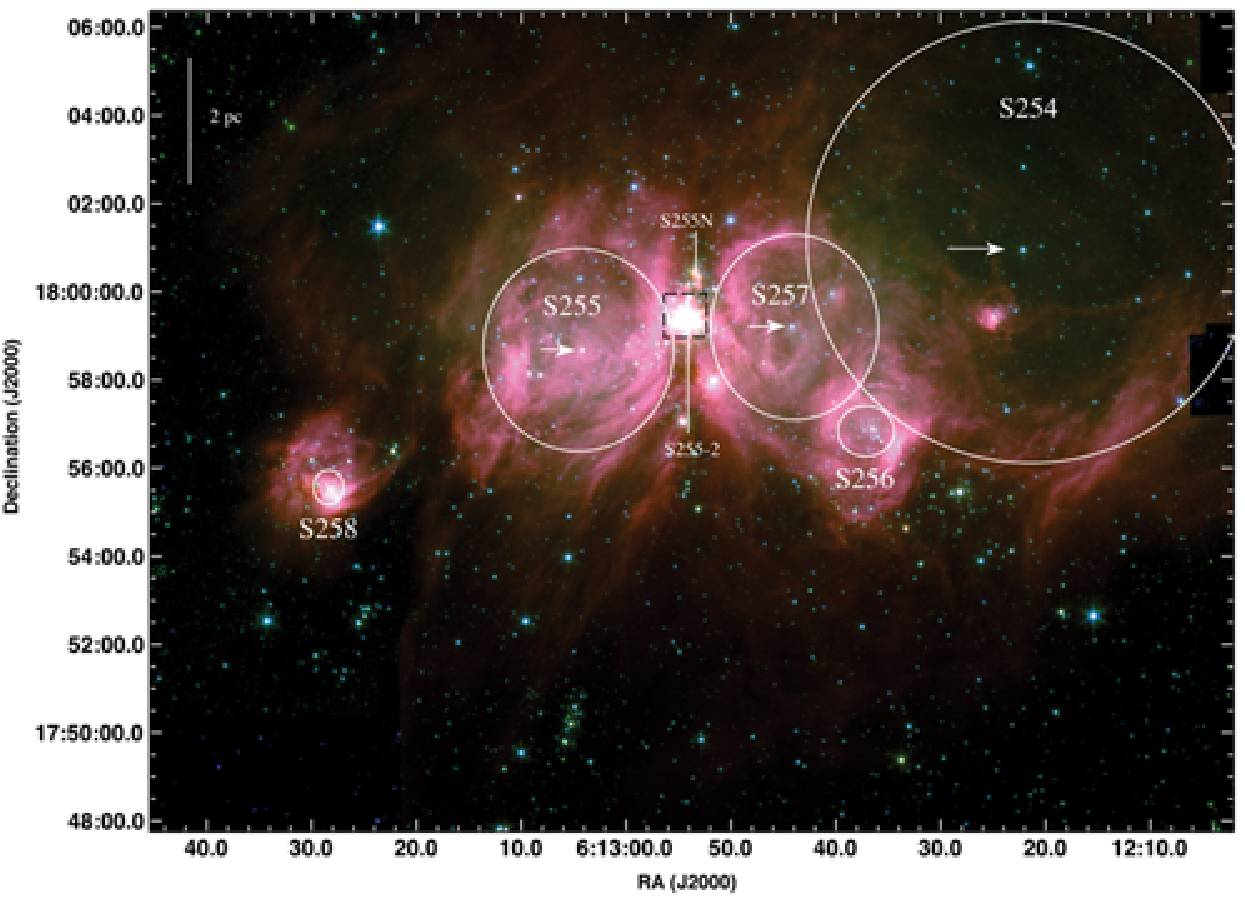}
\caption{IRAC 3-color image of the \complex~complex (blue: 3.6$\mu$m, green: 4.5$\mu$m, red: 8.0$\mu$m). White circles enclose the \hii~regions (Mizuno 1982). Clusters S255-2 and S255N are also indicated. Black rectangle shows the area covered by NIRC data. Known ionizing stars are indicated with arrows.\label{irac124}}
\end{figure}

The S254-S258 complex is a star forming region described for the first time by \citet{sha59} as a group of five \hii~regions (S254 to S258). The complex is located inside an area of 25$\times$20 arc-minutes on the sky (see Figure~\ref{irac124}). The ionizing stars of four of the \hii~regions have been identified: a O9.0V star in S254 \citep{mof79}, B0V in S255 \citep{mof79}, B2.5V in S256 \citep{rus07}, and B0.5V for S257 \citep{ree05}. 

Clusters of stars have been identified in \hii~regions S256, S258, and another inside S254 \citep{dut01}. Until now, there is no reference to the number of cluster members or percentage of sources with IR-excess in those clusters.

A good summary of distance estimations to the complex \complex~can be found in \citet{hun90} and \citet{car95}. In general, there is good agreement on a range of distances between 2 and 3~kpc. We calculated the average of the photometric distances available in the literature and adopted it as the distance to the complex. The photometric distances are 2.6, 2.5, and 2.0 kpc, from \citet{pis76}, \citet{mof79}, and \citet{cha87} respectively, and our adopted distance is 2.4 kpc. This distance is in good agreement with the distances estimated by \citet{eva77} (2-3 kpc, see their appendix), \citet{pis82} (2.3 kpc, kinematic distance using a Schmidt rotation curve), and also \citet{rus07} (2.46 kpc, photometric distance).

Most of the research on the complex has been done on a young massive cluster located between the \hii~regions S255 and S257. This cluster, called S255-2 (also S255IR), is located along a dense and dusty filament of molecular material \citep{hey89}. It contains several signatures of star formation, such as infrared sources, OH, H$_2$O and methanol masers, as well as HH-like objects \citep{bei79,tur71,lo73,min05,mir97}. Approximately one arc-minute north from S255-2, there is a far-IR source detected by \citet{jaf84}, called S255N. This source is also associated with signposts of massive star formation, but since there is no NIR sources associated with S255N, it is believed that this region is in an earlier evolutionary stage than S255-2. \citet{cyg07} show evidence of a massive protocluster in S255N. They found 3 cores at 1.3 mm with no IR counterpart, with masses between 6-35 M$_{\odot}$.\\

\section{Observations and Data Reduction}
\subsection{Mid-IR imaging}
We observed the complex \complex~with the Spitzer space telescope in December 2004 and October 2007. Since the IRAC field of view for the 3.6 and 5.8~$\mu$m bands is shifted with respect to the field of view for 4.5 and 8.0~$\mu$m bands, the area covered by channels 1 and 3 is not the same as the area covered by channels 2 and 4 (see Table~\ref{stars_detected}). The IRAC pixel scale is 1.2 arc-seconds per pixel. We used an integration time of 10.4 seconds per dither in the High Dynamic Range (HDR) mode, with 3 dithers per map position. HDR mode also acquires 0.4 seconds integration time frames for the recovery of bright sources which are saturated in longer exposures. We used S. Carey's artifact correction scripts to remove column pull-down and some of the banding and muxbleed artifacts \citep{hor04}. IRAC mosaics were constructed using the Basic Calibrated Data (BCD) frames (S14.0.0, and S15.3.0) with IRACproc \citep{sch06}. The final image scale is 0.6 arc-seconds per pixel.

We used IRAF DAOPHOT packages to extract sources and perform aperture photometry in each band. The photometry was done using an aperture of 1.8 arc-seconds for the 3.6 and 4.5~$\mu$m bands and 2.4 arc-seconds for the 5.8 and 8.0~$\mu$m bands. Inner and outer sky annuli of 4.8 and 6 arc-seconds respectively were used in each IRAC band. We calculated the zero points in each band using Vega fluxes. The zero points used were 18.443, 17.879, 17.234 and 16.477 for the 3.6, 4.5, 5.8 and 8.0 $\mu$m bands respectively. These include aperture corrections.

\subsection{Near-IR imaging}
Observations with Flamingos were performed at the 2.1 meter telescope located at Kitt Peak National Observatory in December 2004 and January 2006. Flamingos has a $2048\times 2048$ pixel Hawaii II HgCdTe detector array with a plate-scale of 0.611 arc-seconds per pixel which gives a field of view (FOV) of 20$\times$20 arc-min. We obtained imaging at J, H and K bands centered at 1.24 1.65 and 2.21 $\mu$m respectively. The seeing during the observations was approximately 1.0 arc-second. The observations were done in dithering mode with 15-30 arc-seconds shifts for a total integration time of 1000 seconds per band.

The near-IR data reduction was performed using IDL\footnote{Linearization, developed by Robert A. Gutermuth.}, IRAF\footnote{Darks, Flat-field, Bad-pixel mask and background frame creation and application. Distortion correction and mosaicking, developed by Luis A. Chavarr\'{\i}a.} and WCS Tools\footnote{Distortion and astrometry correction, developed by Doug Mink.} routines, which includes linearization, bad-pixel, dark and flat-field creation and application, background frame creation and subtraction, distortion measurements and correction, mosaicking and astrometry correction using the positions of approximately 2000 stars from the 2MASS catalog. Point source detection and aperture photometry were done using IRAF DAOPHOT packages. Aperture photometry for saturated stars was replaced by non-saturated 2MASS photometry and by visual inspection all the detections identified as nebulosity were considered as non-stellar and rejected. Parameters used for detection and aperture photometry were: threshold of 3 sigma for detection and 1.8, 4.8 and 6 arc-seconds in radii for aperture, inner annulus and inner plus outer annulus respectively. Calibration was performed by minimizing residuals to corresponding 2MASS detections. The RMS for residuals between the data sets used was 0.07, 0.09 and 0.08 magnitudes for J, H and K bands respectively.\\

NIRC observations of S255-2 were generously supplied by L. Hillenbrand. The data includes observations in H and K bands, that were taken using the Keck telescope in the year 1999. The resolution is 0.15 arc-seconds per pixel, with a FOV of 1$\times$1 arc-minute. The observations were done in dithering mode with offsets between 3 and 10 arc-seconds for a total integration time of 300 seconds in K-band and 400 seconds in H-band.

We performed aperture photometry of 0.6 arc-seconds in radii for both bands, using a detection threshold of 3 sigma, and 1.5 and 2.25 arc-seconds radii for inner annulus and inner plus outer annulus respectively. As for Flamingos, we calibrated NIRC photometry by minimizing residuals to corresponding 2MASS detections.\\

Finally, we combined IRAC, Flamingos and NIRC data by merging the photometry of the seven bands. In the area covered by both Flamingos and NIRC we used NIRC photometry for all the sources detected in the NIRC mosaics, and Flamingos photometry for sources not detected in the NIRC mosaics. The maximum tolerance in positional offsets between bands was 2.0 arc-seconds.

\subsection{Optical and NIR spectroscopy}
Optical spectra were obtained using Hectospec at the MMT on Mt. Hopkins. We observed the ionizing sources of \hii~regions S254 and S255 \citep{ree05}, and also a source in S256. Near-IR spectrum was obtained for one source in S258 using SpeX \citep{ray03} at the IRTF telescope in Mauna Kea. Hectospec data were reduced using the TDC pipeline at the CfA, and classified using SPTclass~\footnote{http://www.astro.lsa.umich.edu/$\sim$hernandj/SPTclass/sptclass.html}. SpeX data was reduced using Xspextool \citep{cus04}, and telluric correction was performed using Xtellcor \citep{vac03}. The sources in S256 and S258 were chosen because they are located near the center of the \hii~regions, and also because they are the most luminous sources inside the \hii~regions at NIR wavelengths, and are therefore good candidates for being the ionizing source of the \hii~regions (at optical wavelengths, there is one source more luminous than the observed source in S258, but we rejected it because it is located at the edge of the \hii~region).

\subsection{Millimeter}
Observations of the J=1--0 spectral lines of $^{12}$CO and $^{13}$CO were conducted
at the Five College Radio Astronomy Observatory (FCRAO) 14~meter telescope in New Salem,
Massachusetts. The FCRAO beam size is 45 arc-seconds in $^{12}$CO and 
46 arc-seconds in $^{13}$CO. We used the 32 pixel SEQUOIA focal plane array 
\citep{eri99} to image these spectral lines in the on-the-fly (OTF) 
mapping mode. The dual channel correlator allowed the two spectral lines to be 
observed simultaneously, in 1024 channels over a bandwidth of 50~MHz at each frequency. 
The total velocity coverage exceeds 120~km~s$^{-1}$, centered on V$_{\mathrm{LSR}}=0$~km~s$^{-1}$, 
with channel spacings of 0.126~km~s$^{-1}$ ($^{12}$CO) and 0.132~km~s$^{-1}$ ($^{13}$CO).
The velocity resolution is 1.21 times the channel spacing.

These data were taken as part of the Extended Outer Galaxy Survey 
(E-OGS; Brunt \& Heyer, in preparation) that extends the coverage of the
FCRAO Outer Galaxy Survey (OGS; \citet{hey98}) to
Galactic longitude $l$~=~193, over a the latitude range --3.5$\leq$~$b$~$\leq$+5.5.
Further details on the OTF scanning procedure can be found in the E-OGS data paper.
During the survey observations, pointing and focus checks were carried out every 3--4 hours, 
shortly after dawn/dusk or after a significant change in source coordinates. The data
were initially converted to the $T_{A}^{*}$ scale using the standard chopper wheel
method \citep{kut81}. We used the OTFTOOL software, written by M. Heyer,
G. Narayanan, and M. Brewer, to place the spectra on a regular 22.5 arc-seconds grid in
Galactic $l, b$ coordinates. During this procedure, first order baselines were fitted
to the spectra over velocity ranges free of spectral lines.
These baselines were subtracted from the data, and root mean square noise amplitude 
($\sigma$) of each spectrum was recorded. Individual spectra contributing to a single
Galactic coordinate were assigned a $1/\sigma^{2}$ weighting during the gridding. 
The gridded data were scaled to the main beam temperature scale by dividing by the 
main beam efficiency of 0.48.

\section{Results}
The number of sources detected and the total area observed in each band is shown in Table~\ref{stars_detected}. We include only sources with a photometric error of less than 0.2 magnitudes in our study. The 90\% completeness limit in each IRAC and NIR band is also shown in Table 1. Table~\ref{fluxes} lists the coordinates, magnitudes and classification of sources. The complete table is available in the electronic version of this paper. Table~\ref{ionizing_stars} shows the result of our spectral classification for the ionizing stars of the \hii~regions.

\subsection{Description of the IRAC data}\label{section_description}
\begin{figure}
\begin{center}
\epsscale{1.0}
\plottwo{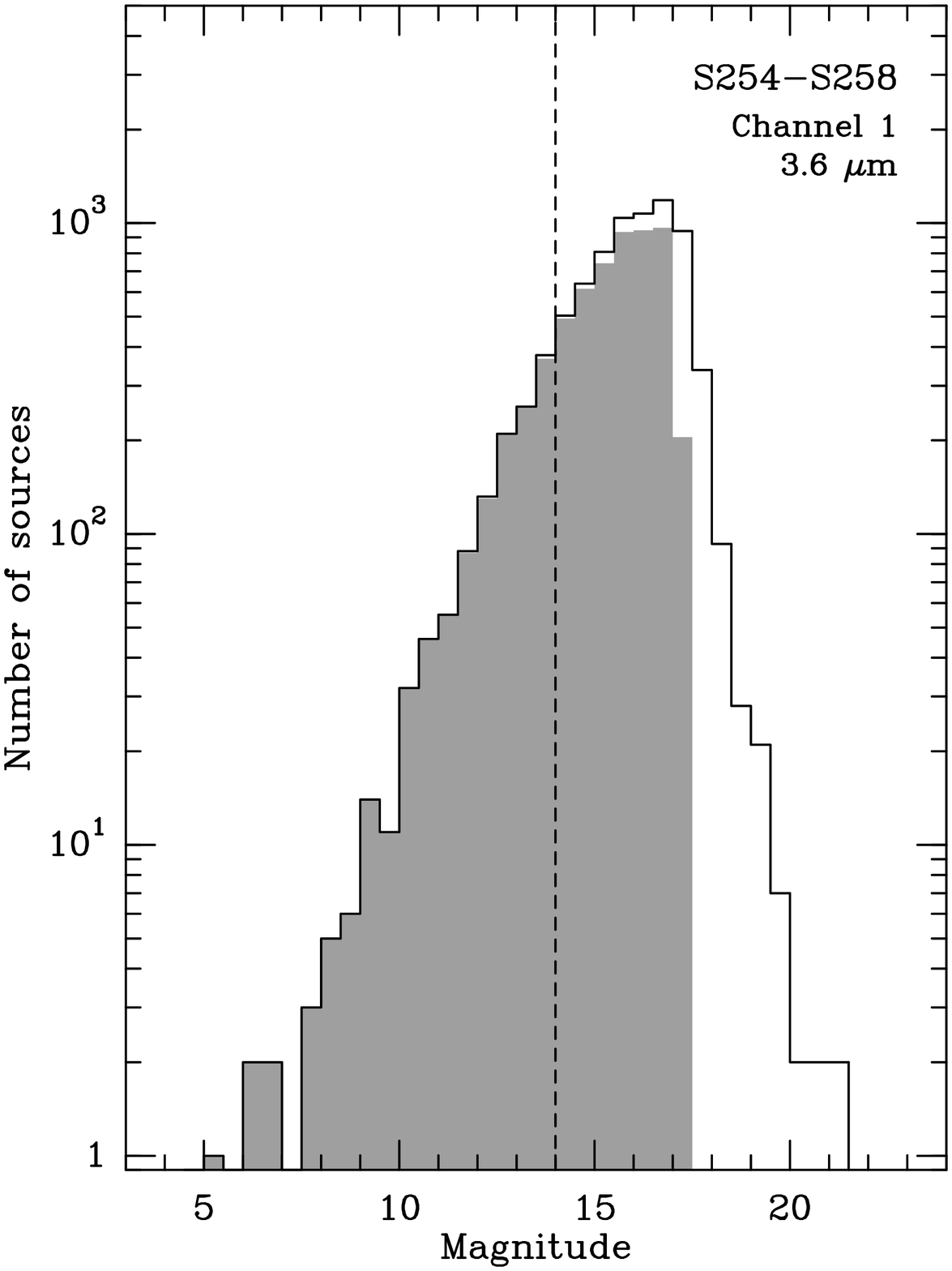}{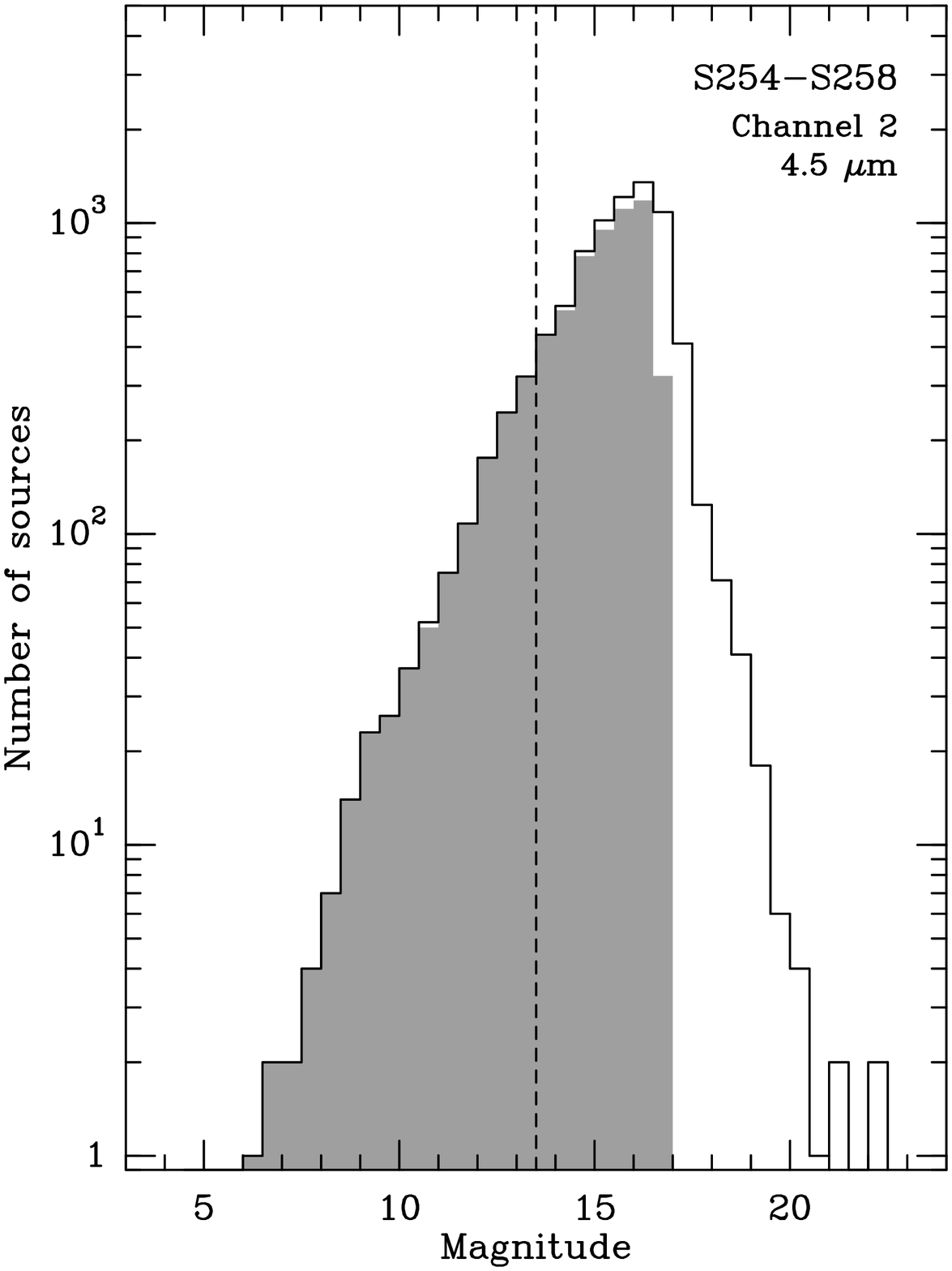}
\plottwo{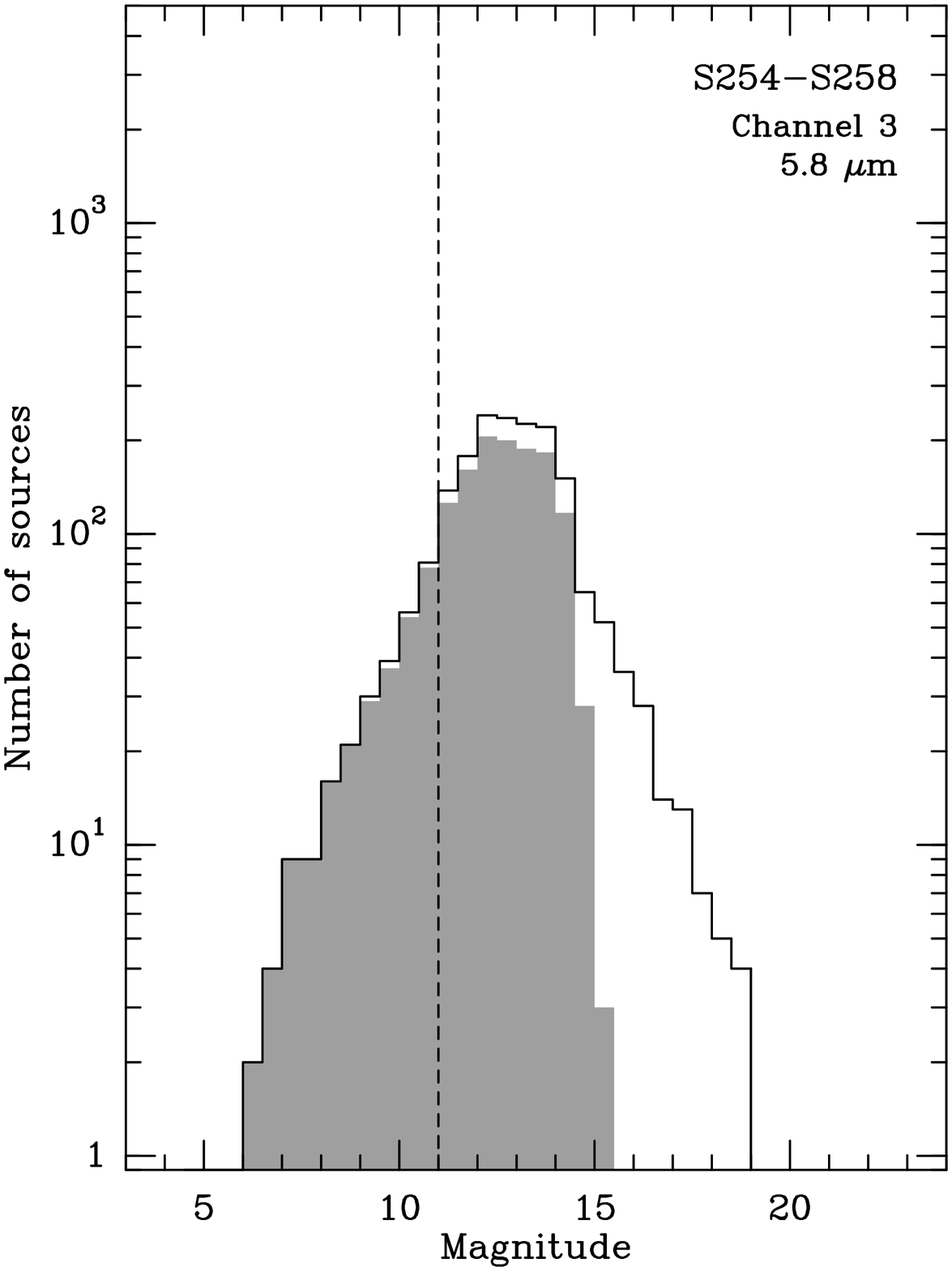}{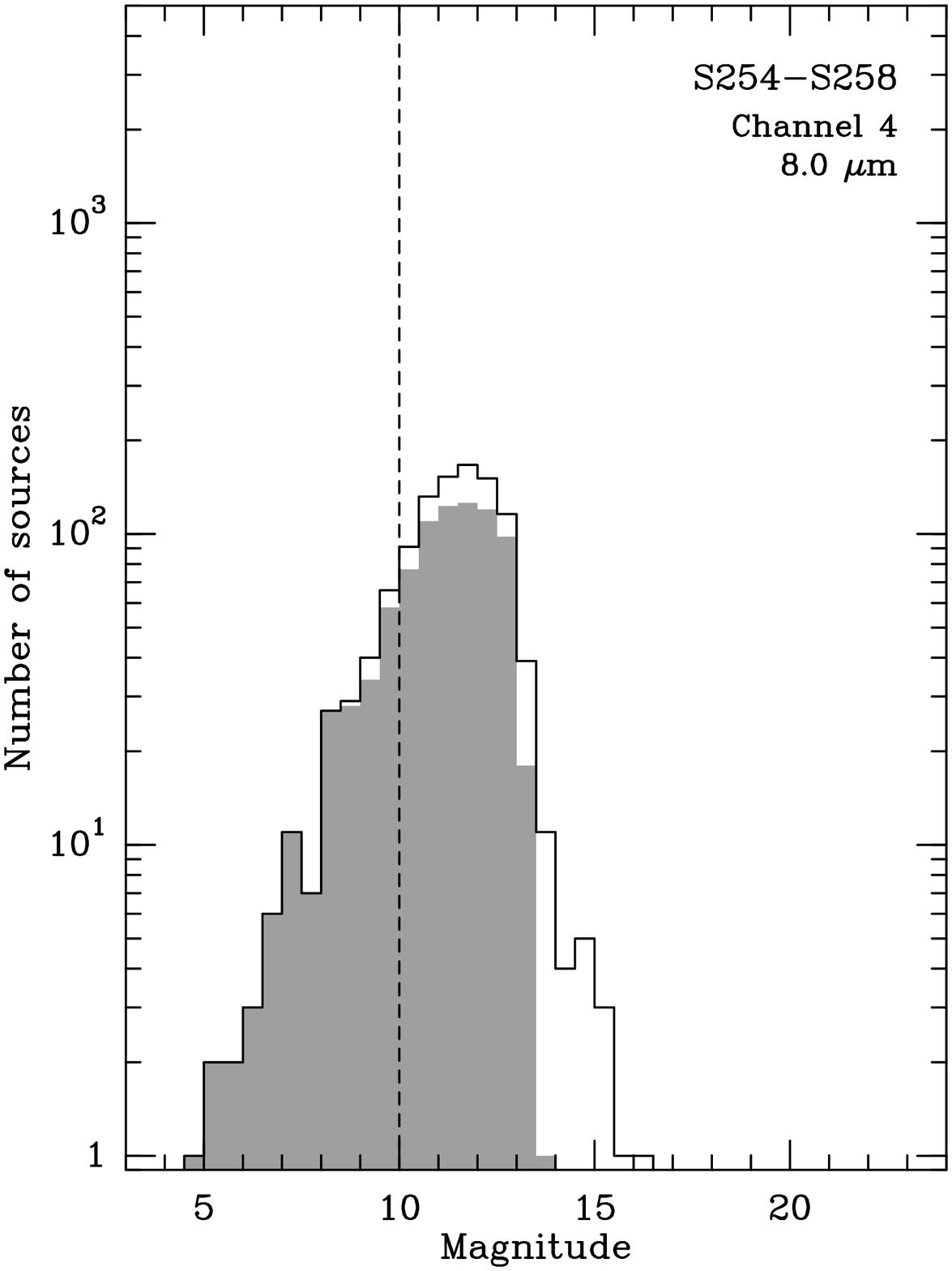}
\caption{Apparent magnitude histograms of detections for all IRAC bands. Dashed area shows detections with error $<$ 0.2 magnitudes. Since 3.6 and 4.5$\mu$m bands are more sensitive to the photospheres of stars than 5.8 and 8.0$\mu$m bands, the number of sources detected at 5.8 and 8.0$\mu$m is less than at 3.6 and 4.5$\mu$m. Dashed lines indicate 90\% completeness. \label{irac_histograms}}
\end{center}
\end{figure}

As shown in Table~\ref{stars_detected}, the number of sources detected with IRAC decreases dramatically in channels 3 and 4 with respect to channels 1 and 2. This difference is caused by the fact that channels 1 and 2 are more sensitive to the photospheric emission from stars. Also, the bright extended emission in channels 3 and 4 decreases our sensitivity (see end of \S~\ref{7band}). The number of sources detected in each of the IRAC bands per magnitude is shown in Figure~\ref{irac_histograms}. We estimate a 90\% completeness at a magnitude of 14, 13.5, 11, and 10 for IRAC channels 1 through 4 respectively. The completeness was estimated by adding and recovering artificial stars. 

\begin{figure}
\begin{center}
\epsscale{1.0}
\plottwo{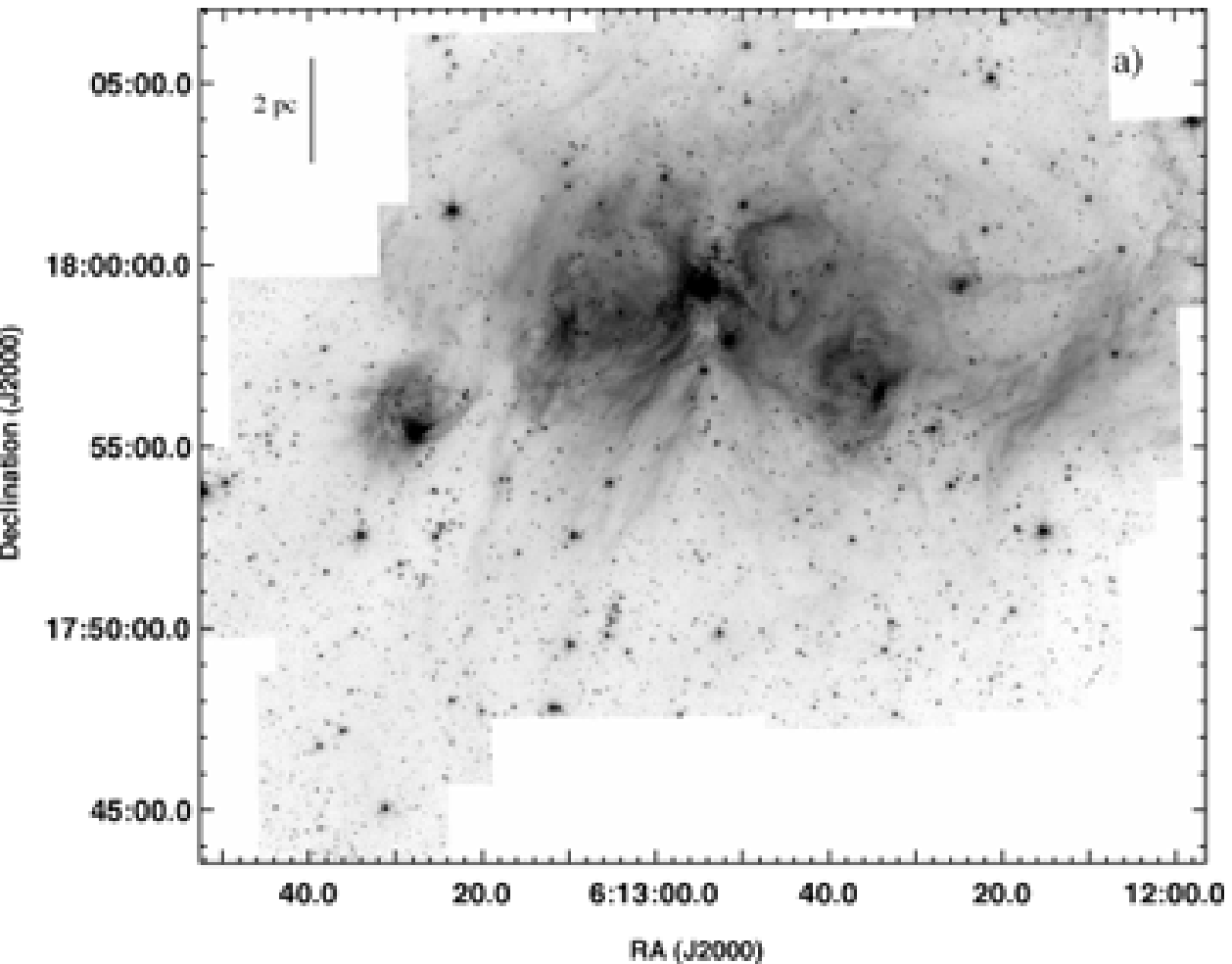}{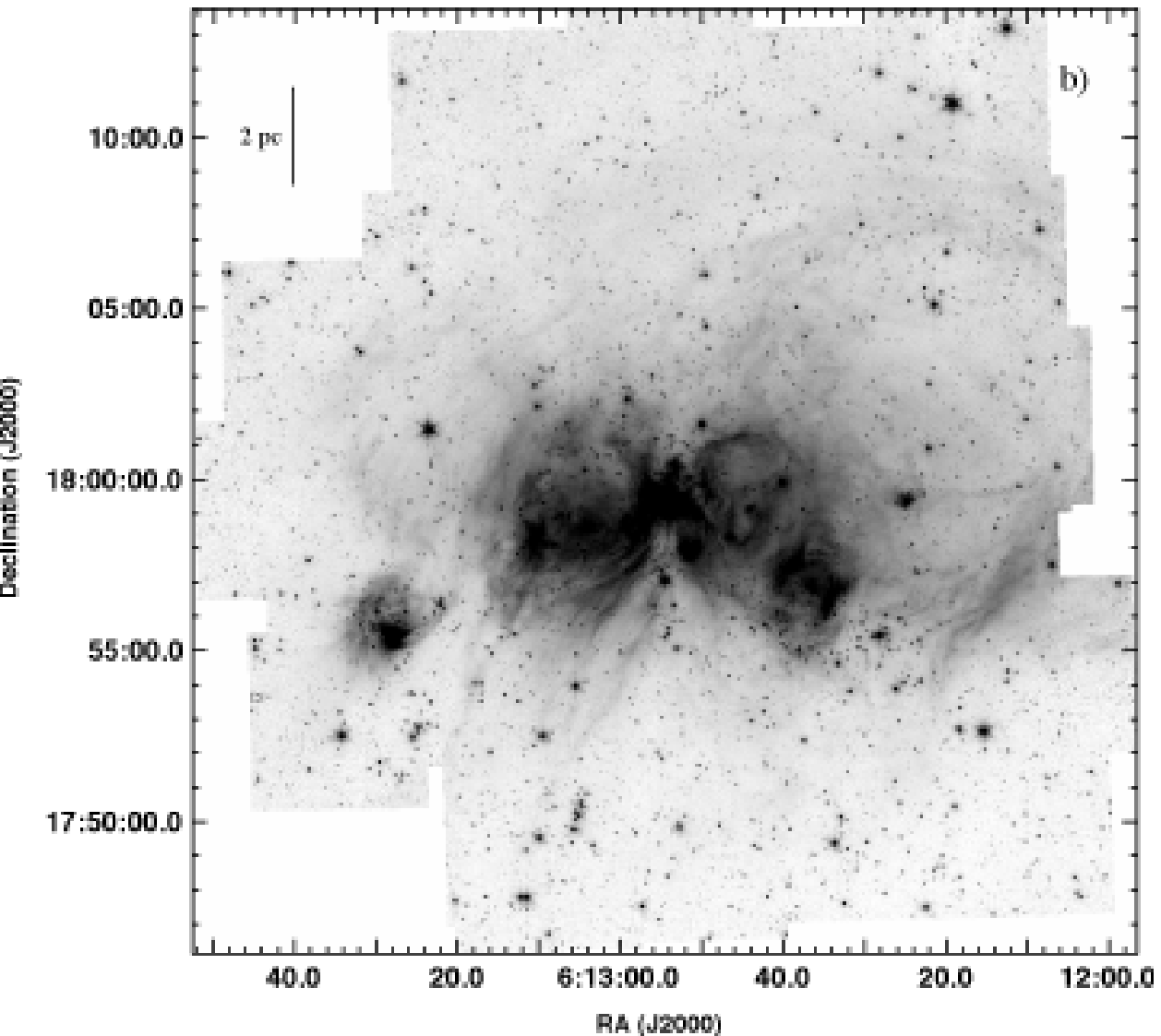}
\plottwo{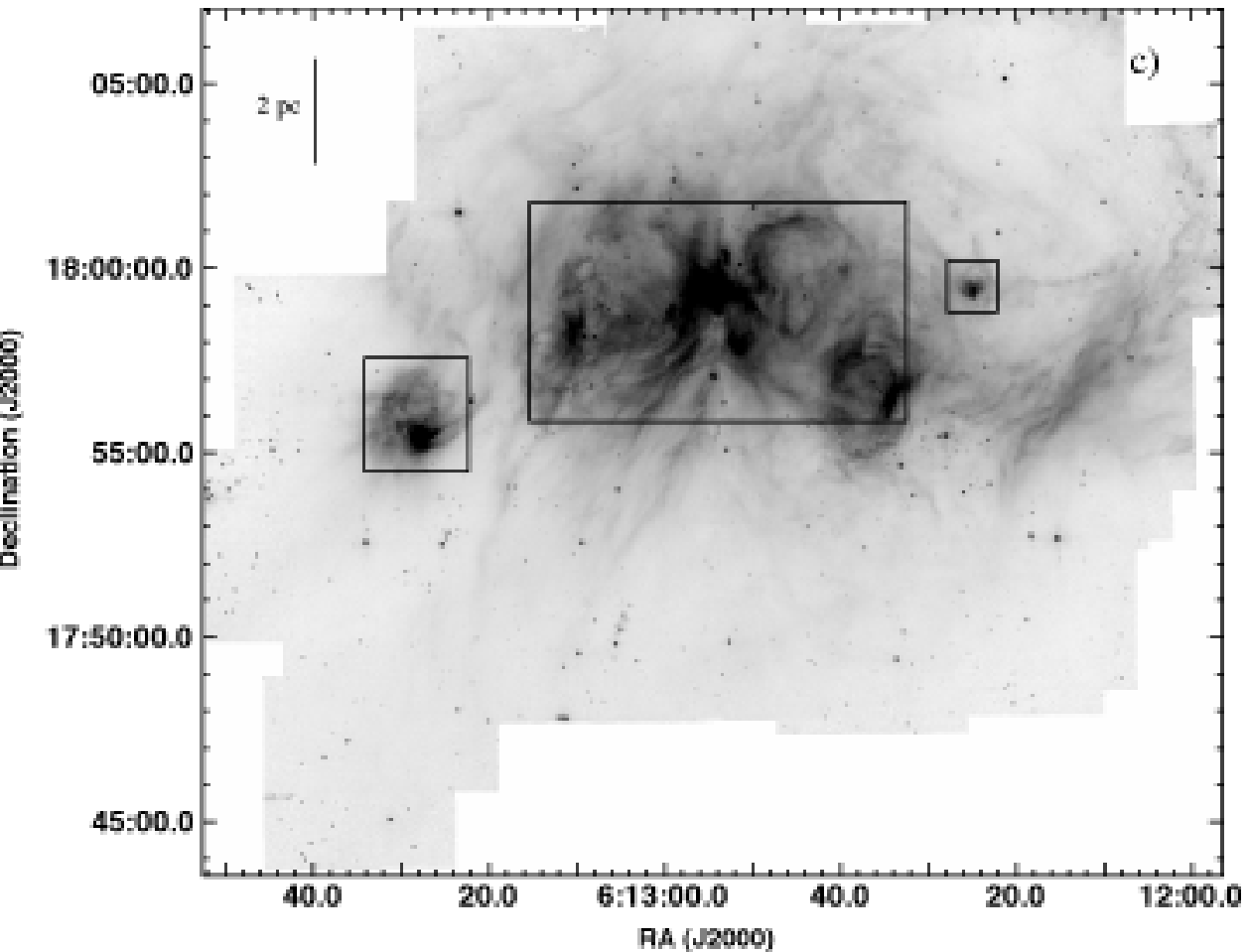}{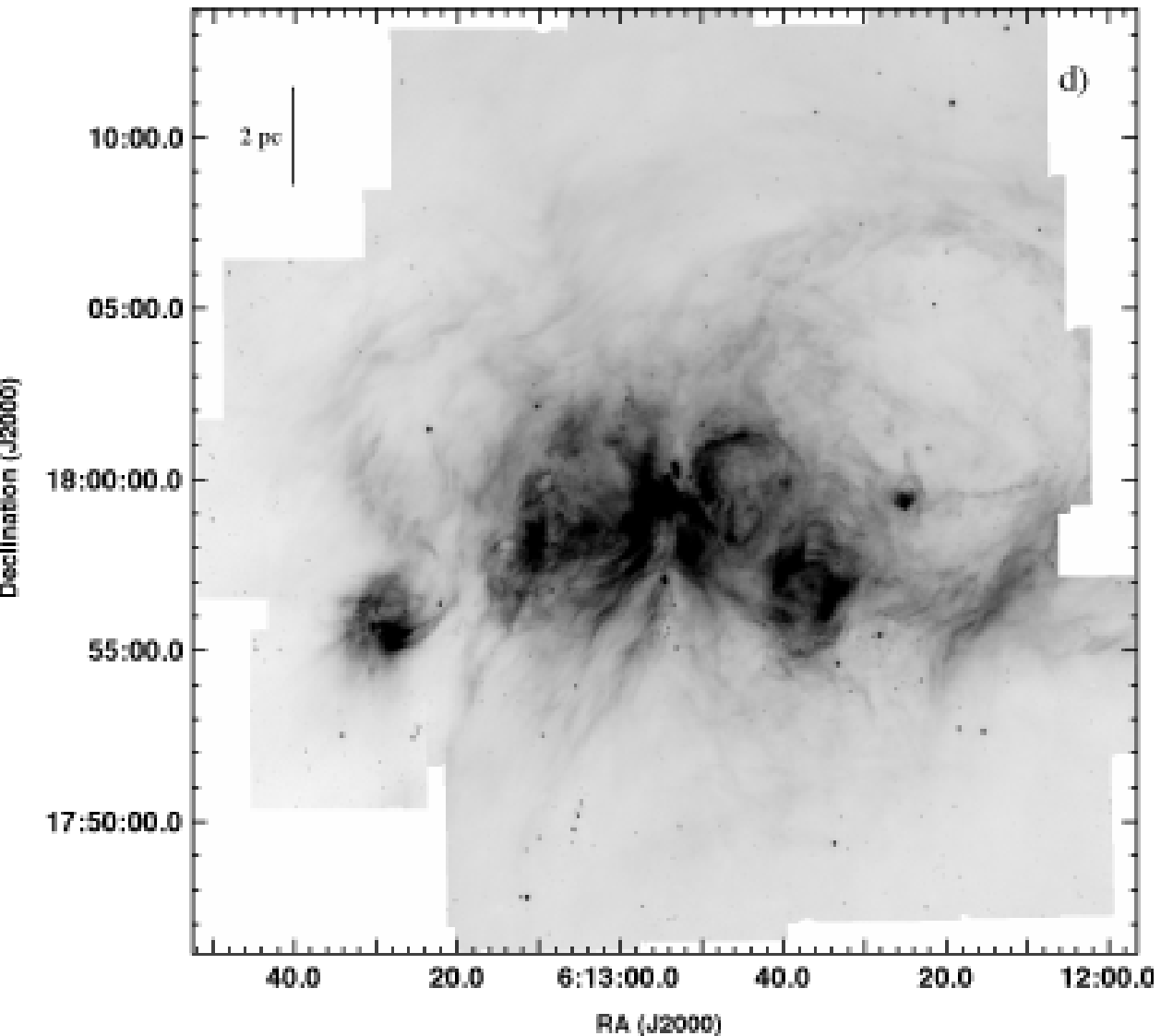}
\caption{Gray-scale IRAC mosaic at: a) 3.6$\mu$m, b) 4.5$\mu$m, c) 5.8$\mu$m and d) 8.0$\mu$m. Boxes in c) enclose areas with bright extended emission. \label{irac_bands}}
\end{center}
\end{figure}

The long exposure mosaic for each of the IRAC channels is shown in Figure~\ref{irac_bands}. Most of the extended emission seen is from polycyclic aromatic hydrocarbon (PAH) features in the IRAC bands (at 3.3, 6.2, 7.7, and 8.6 $\mu$m). The PAH emission is detected around the five \hii~regions in the complex. As shown in the 8 $\mu$m mosaic and in Figure~\ref{irac124}, the PAH emission in S254 forms a shell that roughly follows the perimeter of the \hii~region. The PAH emission inside the shell is very faint, with the exception of the young cluster located south-east from the ionizing star. The PAH emission in S255 and S257 has a similar morphology as in S254, with more PAH emission coming from the inner parts of the \hii~regions. In S256 the PAH emission has a shell-like morphology with emission from the inner part of the \hii~region. The structure of the PAH emission in S258 has no shell-like structure visible.

\subsection{Identifying and classifying young stars}\label{section_classification}
In this section, we describe how YSO were identified and classified. We classified the YSO using two methods: a) the observed IRAC spectral energy distribution (SED) slope (\ai), for sources detected in all four IRAC bands, and b) the dereddened IRAC SED slope (\ai), for sources with 7-band detections (J, H, K, and the four IRAC bands). We identified additional YSO by combining the H, K and IRAC 4.5 $\mu$m bands. Probable background sources were eliminated using the algorithms explained in \S~\ref{section_background}.

\subsubsection{Sources with IRAC 4-band detections}
The SED slope is defined as:
\be
\alpha = \frac{dlog(\lambda F_{\lambda})}{dlog(\lambda)},
\ee
and was introduced as a tool to classify YSO by \citet{lad87}. In that paper, the YSO were classified in three categories; Class I (star plus infalling envelope), Class II (star plus a disk), and Class III (post T-tauri stars). The classification of YSO evolved with time, adding more categories like Class 0 \citep{and00}, or flat spectrum \citep{gre94}, as well as changing the boundaries between them.

\begin{figure}
\epsscale{0.6}
\plotone{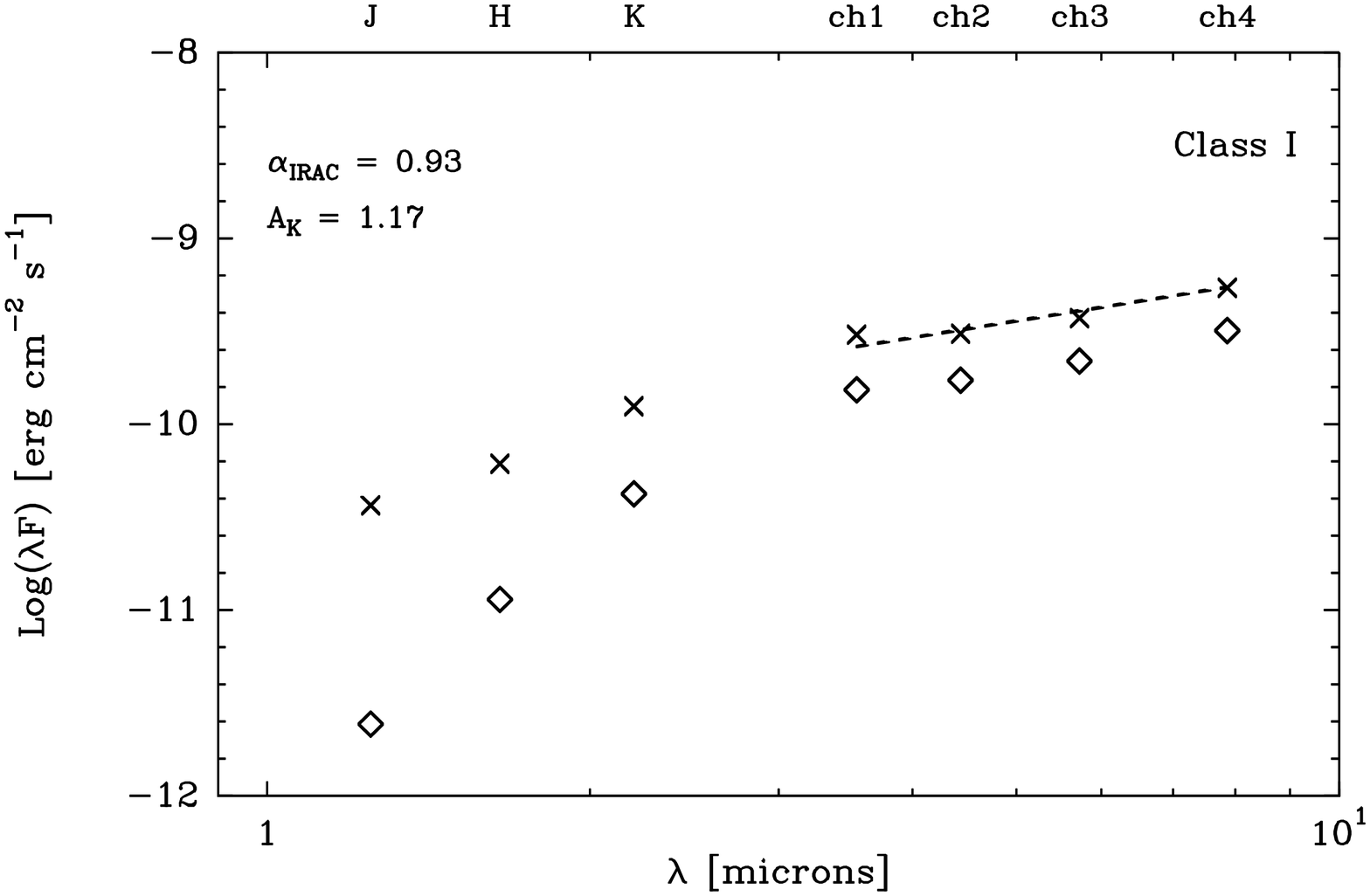}
\plotone{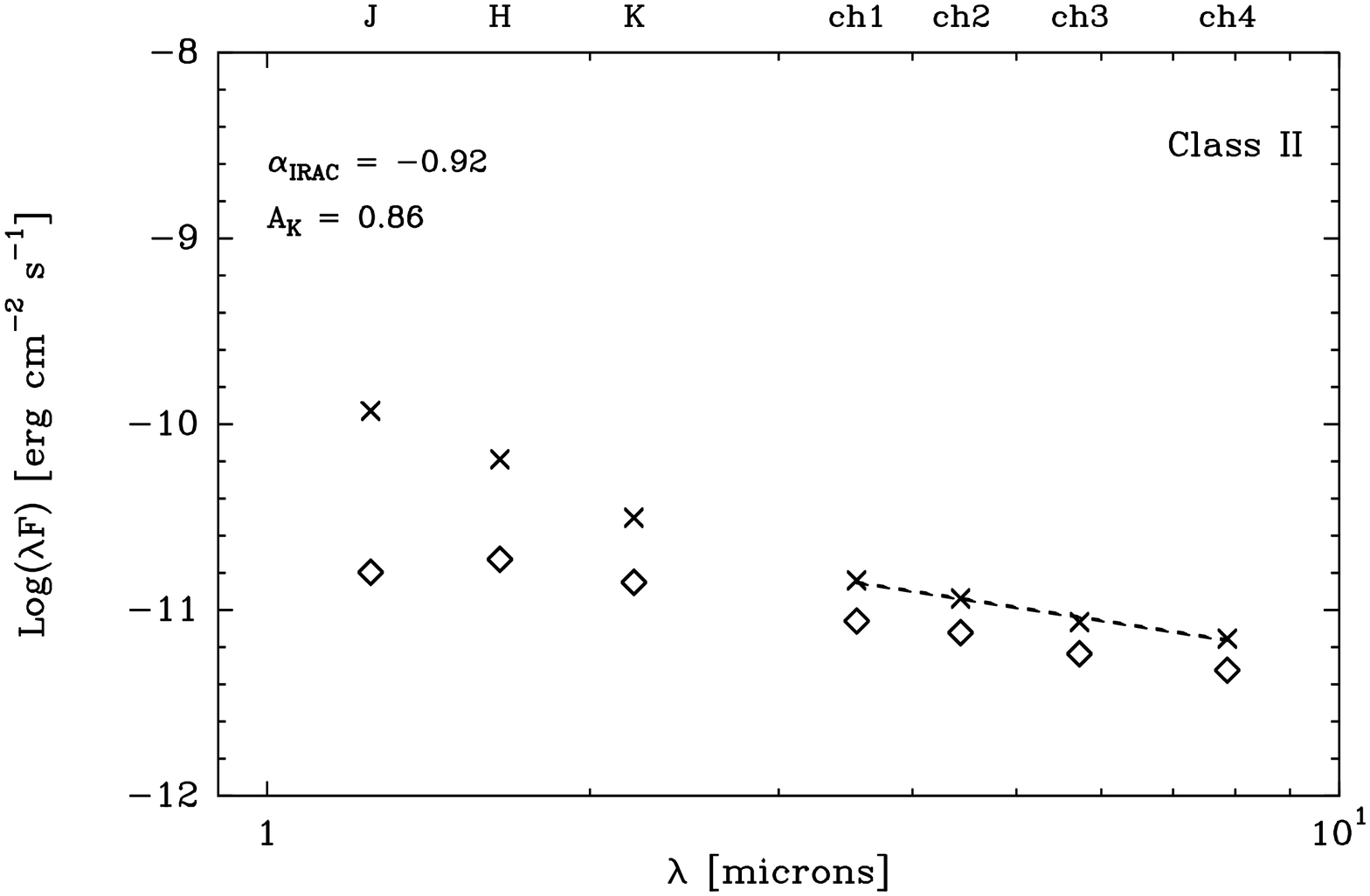}
\caption{Spectral energy distribution (SED) of a Class I (\emph{left}) and Class II (\emph{right}) source in the \complex~complex. The observed fluxes in the JHK and all IRAC bands are shown by diamonds. Dereddened fluxes are shown by crosses. Dashed line shows the fitted \ai.\label{irac_sed}}
\end{figure}

Here, we use the IRAC SED slope (\ai) to classify the YSO using the following criteria: Class I ($0 <$ \ai), Class II ($-2 \leq$ \ai~$\leq 0$), and Class III or main sequence photospheres (\ai~$< -2$). There could be some Class 0 sources in our Class I category, but we are not able to separate them. This analysis was done for all the sources that were detected in all four bands. We use \ai~because the slope of the SED is less affected by extinction in the IRAC wavelengths than in the near-IR. To estimate \ai, we performed a minimum $\chi^2$ fit over the IRAC fluxes (see Figure~\ref{irac_sed}).

\begin{figure}
\epsscale{0.6}
\plotone{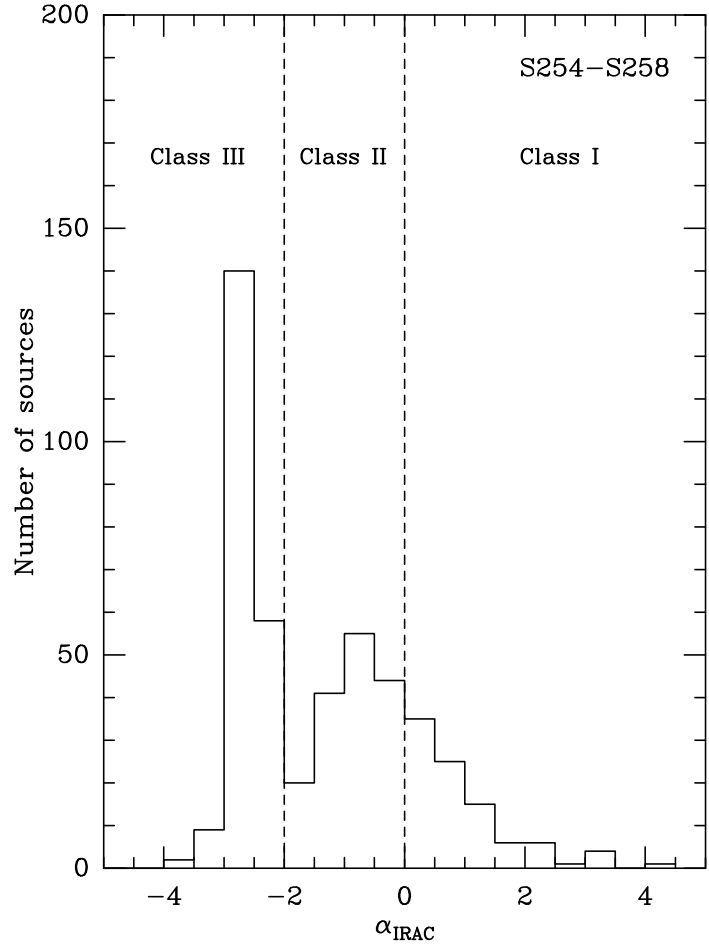}
\caption{Distribution of spectral indeces of sources detected in all IRAC bands. Dashed lines separate Class I and Class II sources.\label{spectral_histograms}}
\end{figure}

Figure~\ref{spectral_histograms} shows the distribution of the observed \ai~in the \complex~complex. The distribution is similar to the distribution seen in nearby star forming regions \citep{lad06,kum07,har07}, and also in the SED models by \citet{rob06}. The peak corresponding to Class II sources may be explained by the amount of time that stars stay in a given stage of evolution. 

We detected a total of 462 sources in all four IRAC bands. Of those, using the observed \ai, 93 were classified as Class I and 160 as Class II.

\subsubsection{Sources with 7-band detections}\label{7band}
For sources that have NIR detections in J, H, and K-bands in addition to the four IRAC bands, we can correct for extinction and use the dereddened value of \ai~to classify them. The sources were dereddened by moving them to their intrinsic main sequence color in the H-K vs. J-H color-color diagram \citep{bes88} using the reddening law from \citet{ind05} in JHK-bands. After the sources were dereddened, we performed a minimum $\chi^2$ fit over the dereddened fluxes in the IRAC wavelengths to estimate \ai, and then we classified them using the criteria explained in the previous section. We used the extinction law from \citet{fla07} to deredden the sources in the IRAC wavelengths.

We detected a total of 264 sources in all 7-bands. Of those, using the dereddened \ai, 26 were classified as Class I and 109 as Class II. If we compare this result with the classification of the same sources using the observed \ai, we find that 17 (6\%) change their classification: 6 Class I became Class II, and 11 Class II became Class III after correcting for extinction. 

\begin{figure}
\epsscale{0.6}
\plotone{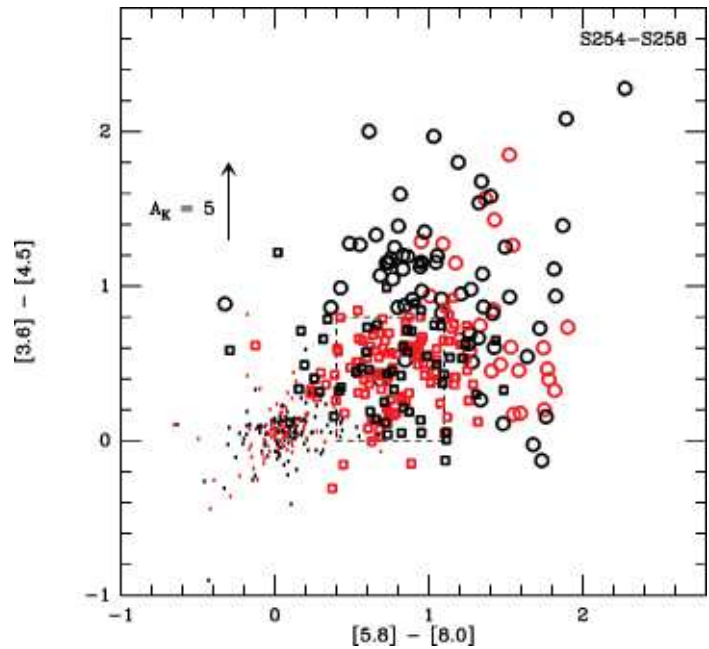}
\caption{IRAC color-color diagram of observed magnitudes for all sources detected in all four IRAC bands. Sources detected only in the IRAC bands (shown in black) were classified using their observed \ai~value as Class I ($\alpha_{IRAC} > 0$, shown as circles), Class II ($-2 \leq \alpha_{IRAC} \leq 0$, shown as squares), or Class III ($\alpha_{IRAC} < -2$, shown as points). Sources that were also detected in JHK (shown in red), were classified using their dereddened \ai~value. Large dashed rectangle indicates the position of Class II sources from Allen et al. 2004. The reddening vector is from Flaherty et al. 2007.\label{irac_colcol}}
\end{figure}

Finally, by combining the two methods, that is, using the 4-band classification method and correcting by the dereddened value of \ai~in sources with 7-band detections, we detected in total 87 Class I, and 165 Class II out of 462 sources in the \complex~complex. Figure~\ref{irac_colcol} shows an IRAC color-color diagram with the sources coded by classification.

\begin{figure}
\begin{center}
\epsscale{1.0}
\plottwo{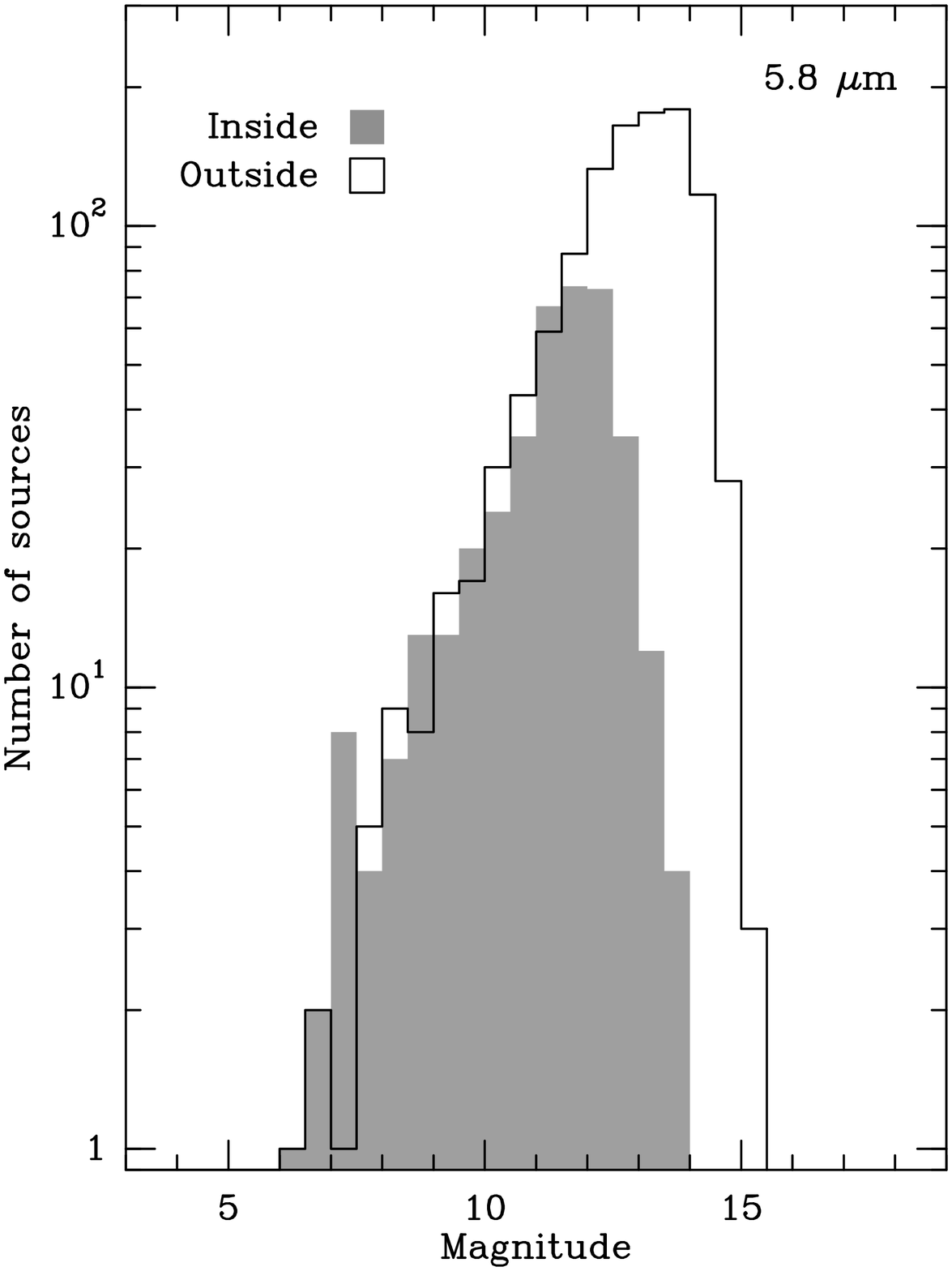}{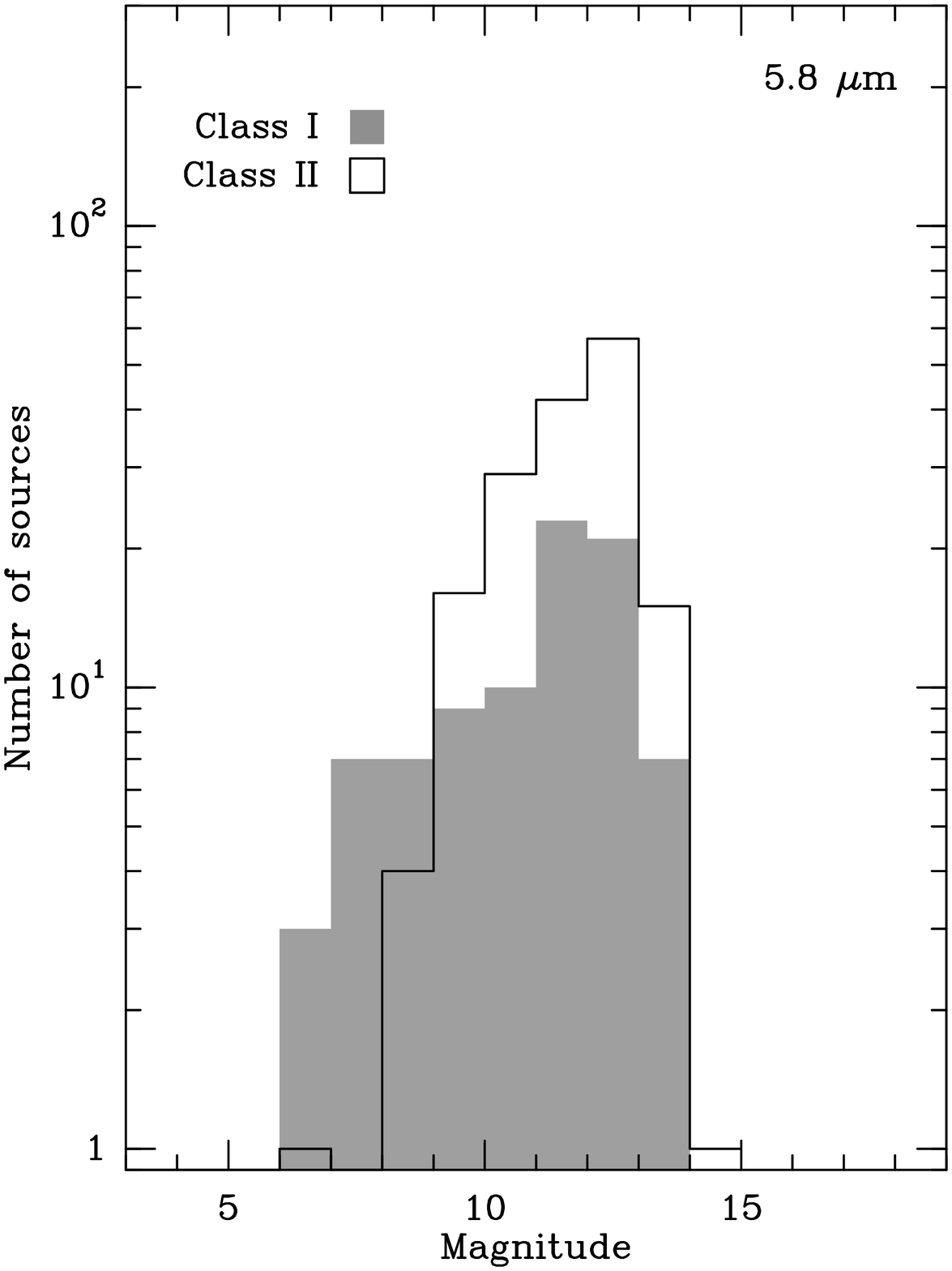}
\plottwo{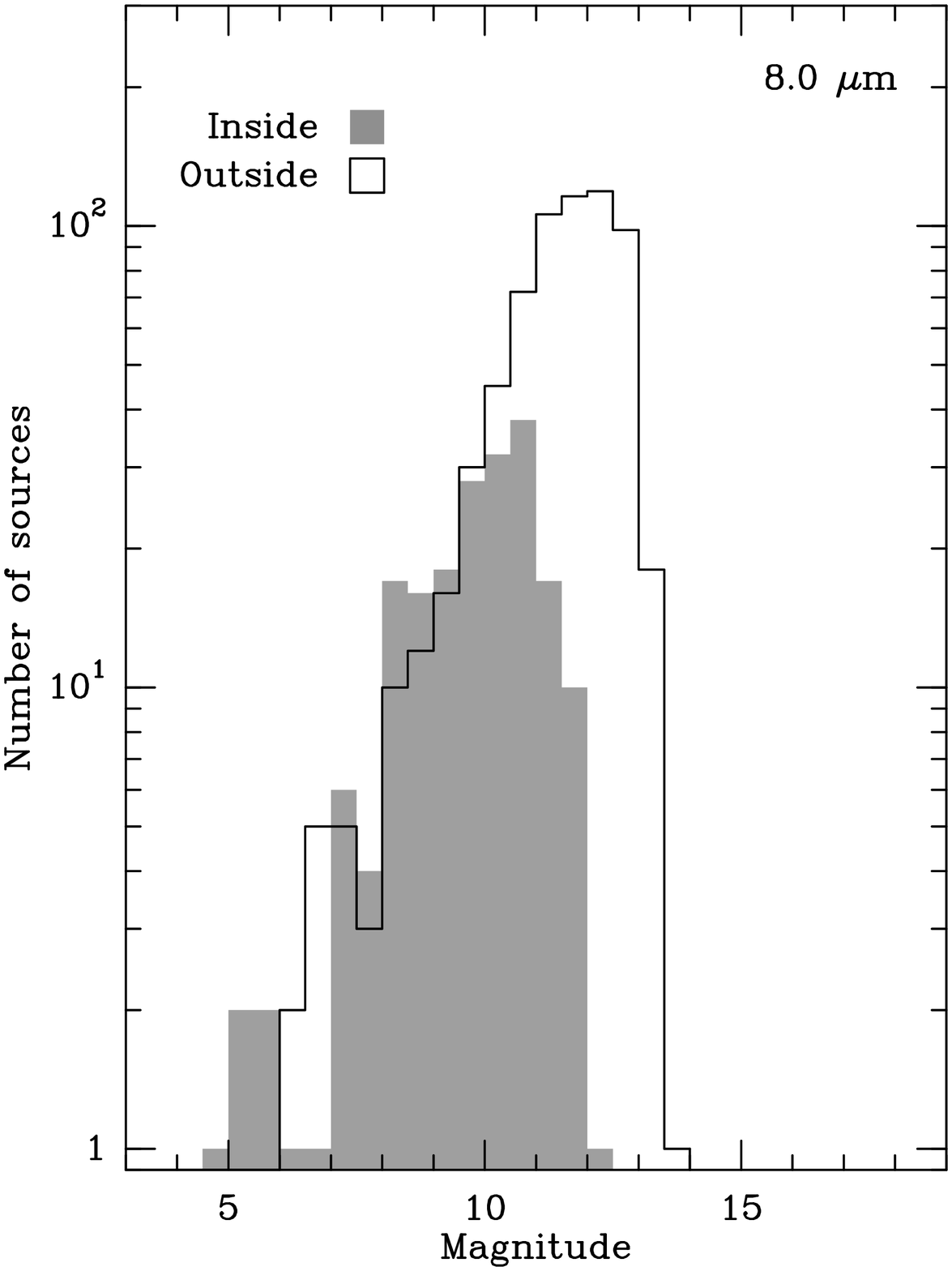}{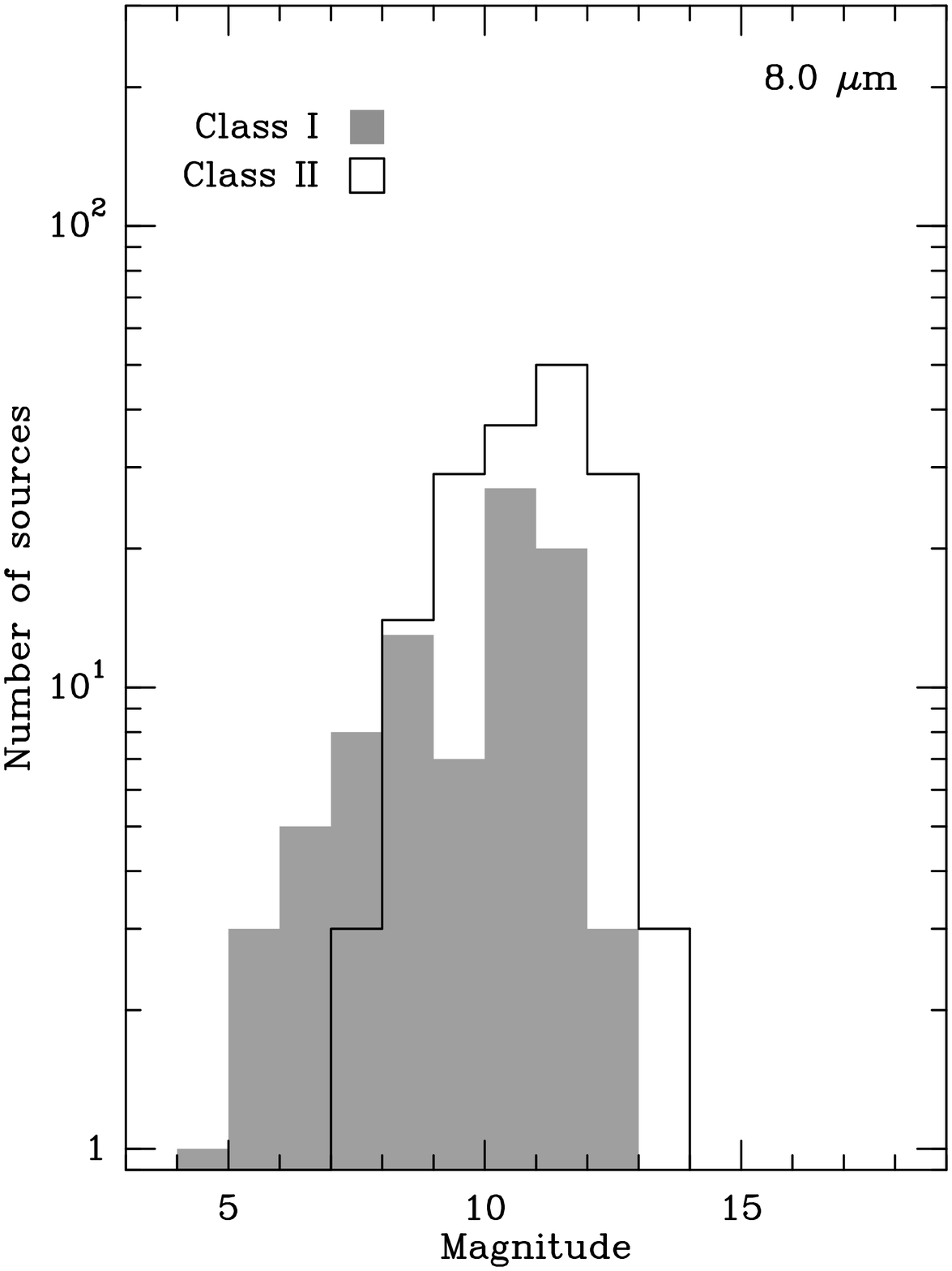}
\caption{\textbf{Upper-left}: Distribution of sources detected inside (gray), and outside (white) areas with bright PAH emission (see areas marked in Figure~\ref{irac_bands}c) at 5.8 $\mu$m. \textbf{Upper-right}: Distribution of Class I (gray), and Class II sources at 5.8 $\mu$m. \textbf{Lower}: The same but at 8.0 $\mu$m. We don't detect faint sources inside the area of bright PAH emission.\label{pah_ch3}}
\end{center}
\end{figure}

\begin{figure}
\epsscale{0.6}
\plotone{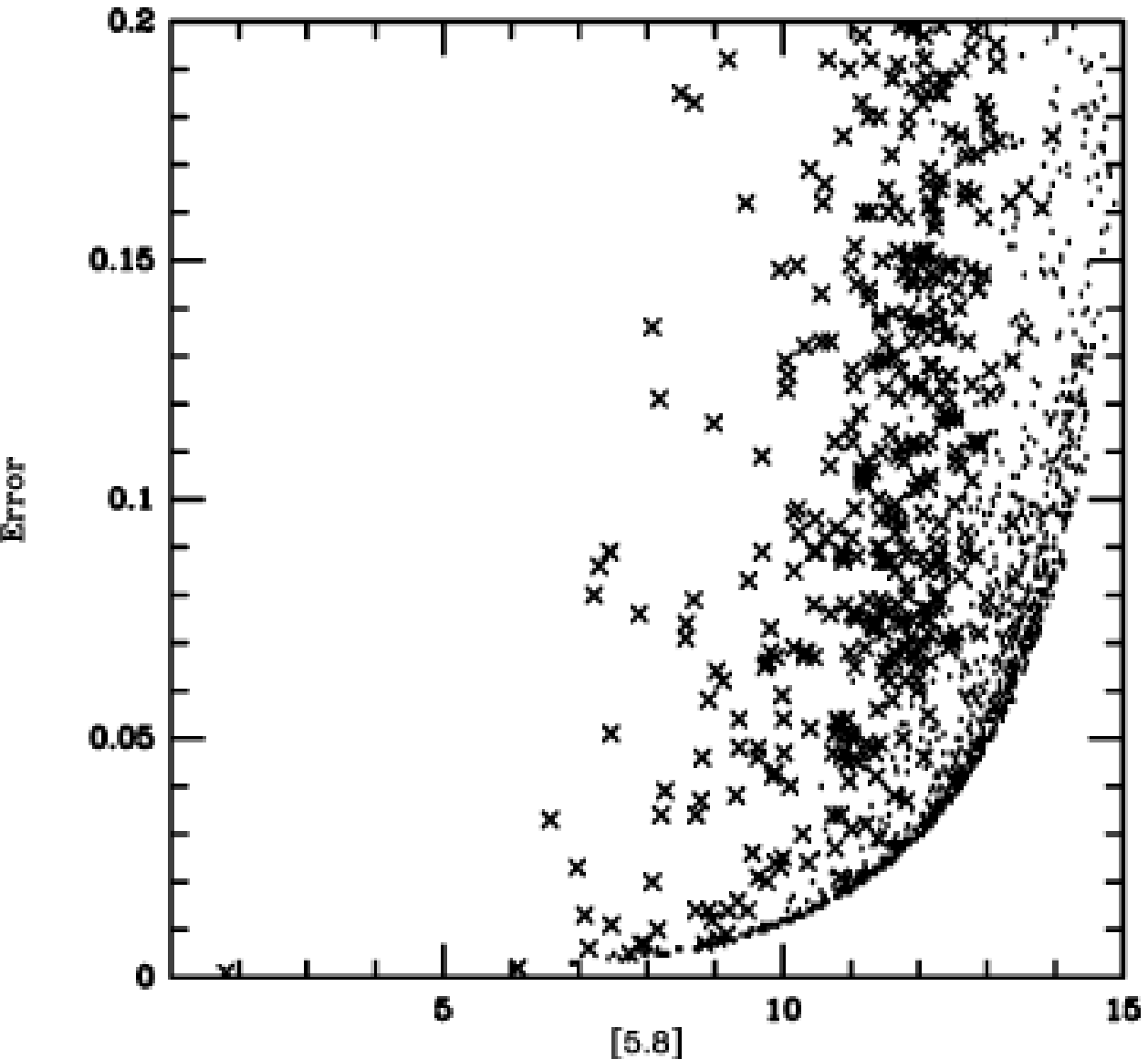}
\plotone{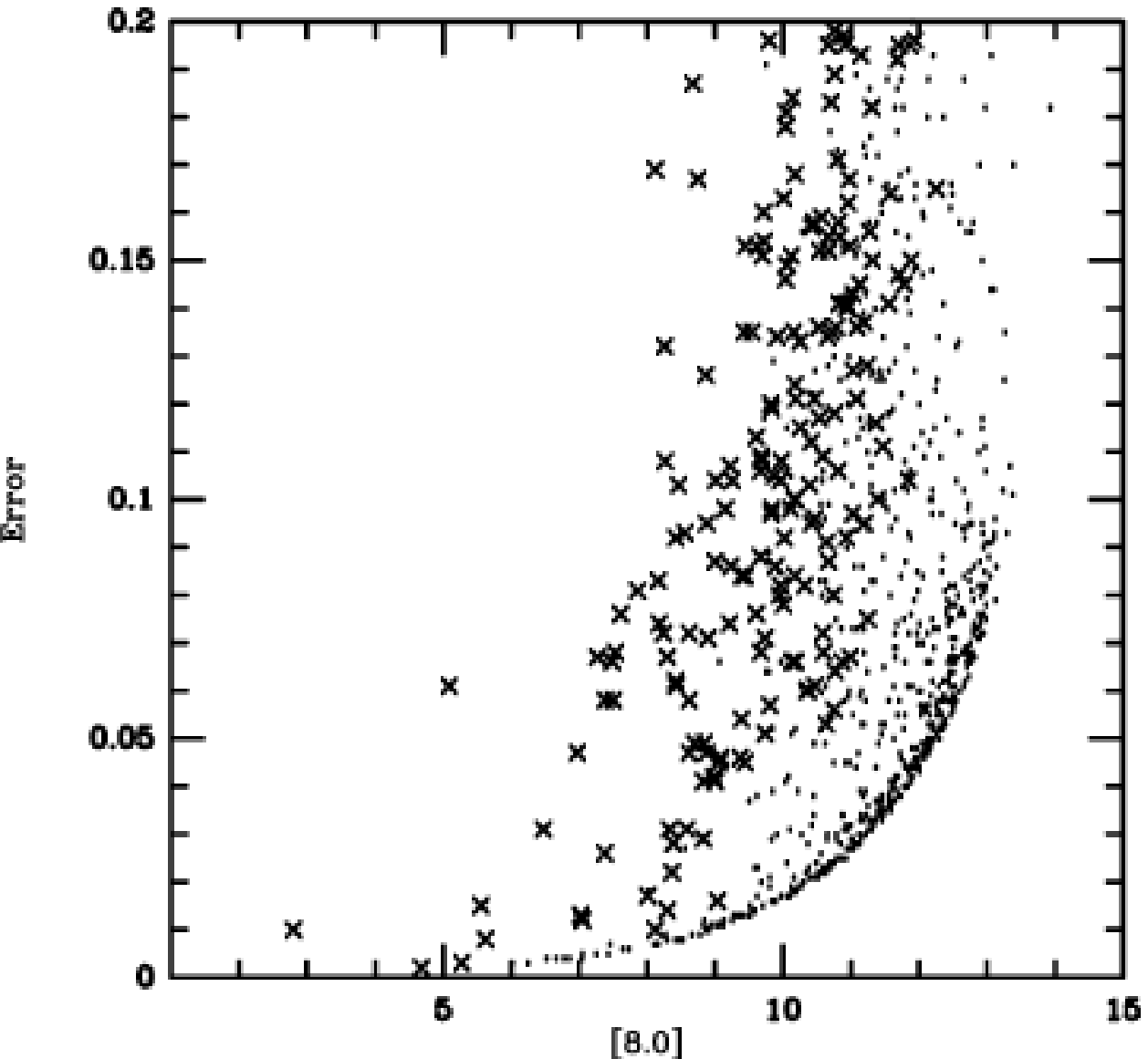}
\caption{Photometric uncertainty of sources detected at 5.8~$\mu$m (\emph{upper}) and 8.0~$\mu$m (\emph{lower}). Sources located inside areas with bright PAH emission are marked as crosses, sources located outside those areas are marked as points. Sources located outside areas with bright PAH emission extend to fainter magnitudes.\label{pah_error}}
\end{figure}

The PAH emission present in the complex decreases the sensitivity to faint sources at 5.8 and 8.0 $\mu$m in areas where the emission is bright. In order to quantify this effect, we compared the magnitudes of sources located in areas with bright PAH emission against sources located in areas with faint PAH emission. Figure~\ref{pah_ch3} shows the distribution of YSO detected inside areas with bright emission (in gray), and outside those areas (white) at 5.6 and 8.0 $\mu$m (areas with bright emission are defined in Fig.~\ref{irac_bands}c). Also, it shows the distribution of Class I (in gray) and Class II (in white) sources detected at 5.6 and 8.0 $\mu$m. From the histograms we can deduce: 1) the sources detected outside the areas with bright PAH emission extend to fainter magnitudes than inside those areas (this can also be seen in Figure~\ref{pah_error}), and 2) the distribution of Class I and Class II shows that on average Class II sources are fainter than Class I sources. We conclude that we are not detecting some of the faint YSO in locations with bright PAH emission, and the undetected YSO are preferentially Class II. The regions that are going to be likely affected by this bias are located between \hii~regions S255 and S257.

\subsubsection{Identifying more YSO} \label{IRexcess}
We used a combination of H, K, and IRAC 4.5 $\mu$m bands to detect more sources with IR-excess. We used the 4.5 $\mu$m band because it is more sensitive to IR excess than 3.6 $\mu$m. Since now we do not require detections in IRAC bands 3 and 4, we are able to detect faint sources not detected before due to the bright extended emission in those channels.

\begin{figure}
\epsscale{0.6}
\plotone{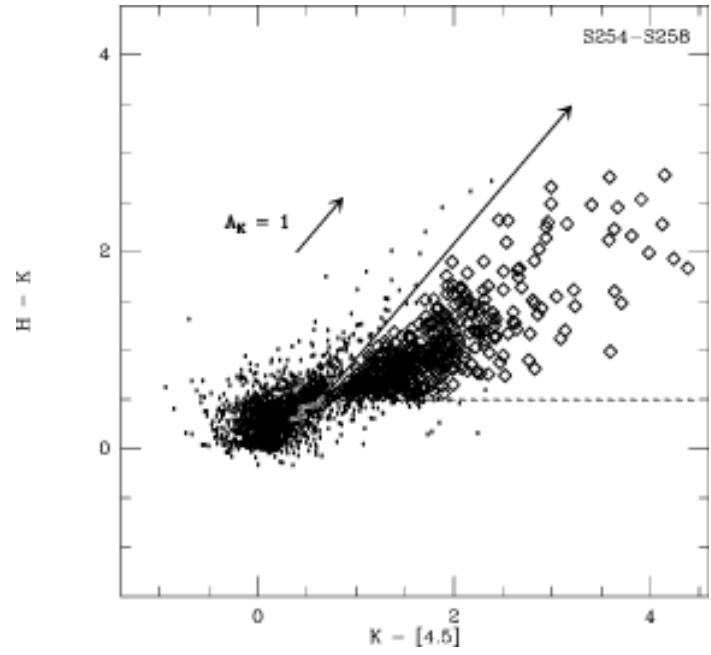}
\caption{Flamingos H, K and IRAC 4.5 $\mu$m band color-color diagram for \complex. The black line near (0,0) shows the main sequence locus of a late type M dwarf star \citep{pat06}. The long arrow corresponds to the reddening vector (from Flaherty et al. 2007). Stars with infrared excess, shown as diamonds, lay between the reddening vector and the dashed horizontal line.\label{flmn_irac_colcol}}
\end{figure}

Figure~\ref{flmn_irac_colcol} shows the IRAC$+$NIR color-color diagram using H, K and 4.5 $\mu$m. In the diagram, the slope of the reddening vector was determined using the extinction law from \citet{fla07}. The stars with infrared excess are located to the right of the reddening vector and plotted as diamonds in the diagram. In order to classify a source as having an IR-excess, the source must be located more than 1 sigma away from the reddening vector.

In total, we found 395 stars with NIR excess with this method. Of those, 138 correspond to sources detected in the 4 IRAC bands; 38 Class I, 99 Class II, and 1 Class III.\\

Adding all the sources with NIR excess and the Class~I and Class~II sources described in previous sections, we detected in total 510 sources with IR-excess in the \complex~complex.

\subsubsection{Background contamination} \label{section_background}
Due to the sensitivity of IRAC observations and the relative transparency of molecular clouds at mid-infrared wavelengths, we expect to have field contamination in our sample. Results from the Spitzer c2d Survey have shown that galaxies have colors that are similar to the colors of young stars \citep{jor06}. In order to separate those galaxies from the YSO, we used a color-magnitude analysis to isolate most ($>$ 90$\%$) of the background contamination.

The data used in our background analysis were taken from the IRAC Shallow Survey \citep{eis04}. This survey covers 8.5 square degrees of the sky in the NOAO Deep Wide-Field Survey in Bo\"otes with 30 second exposures, which makes it deeper than our IRAC data.

\begin{figure}
\epsscale{0.6}
\plotone{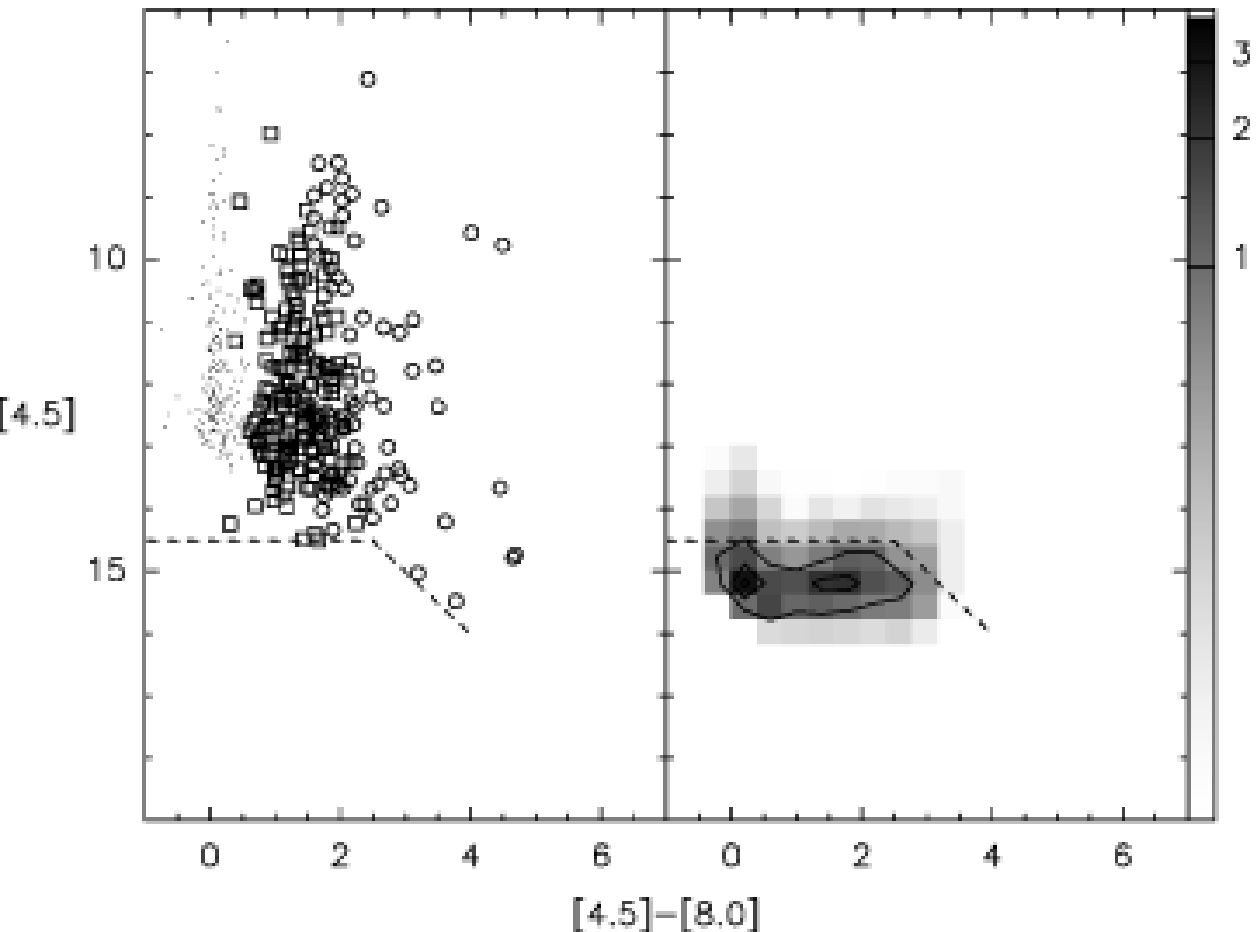}
\plotone{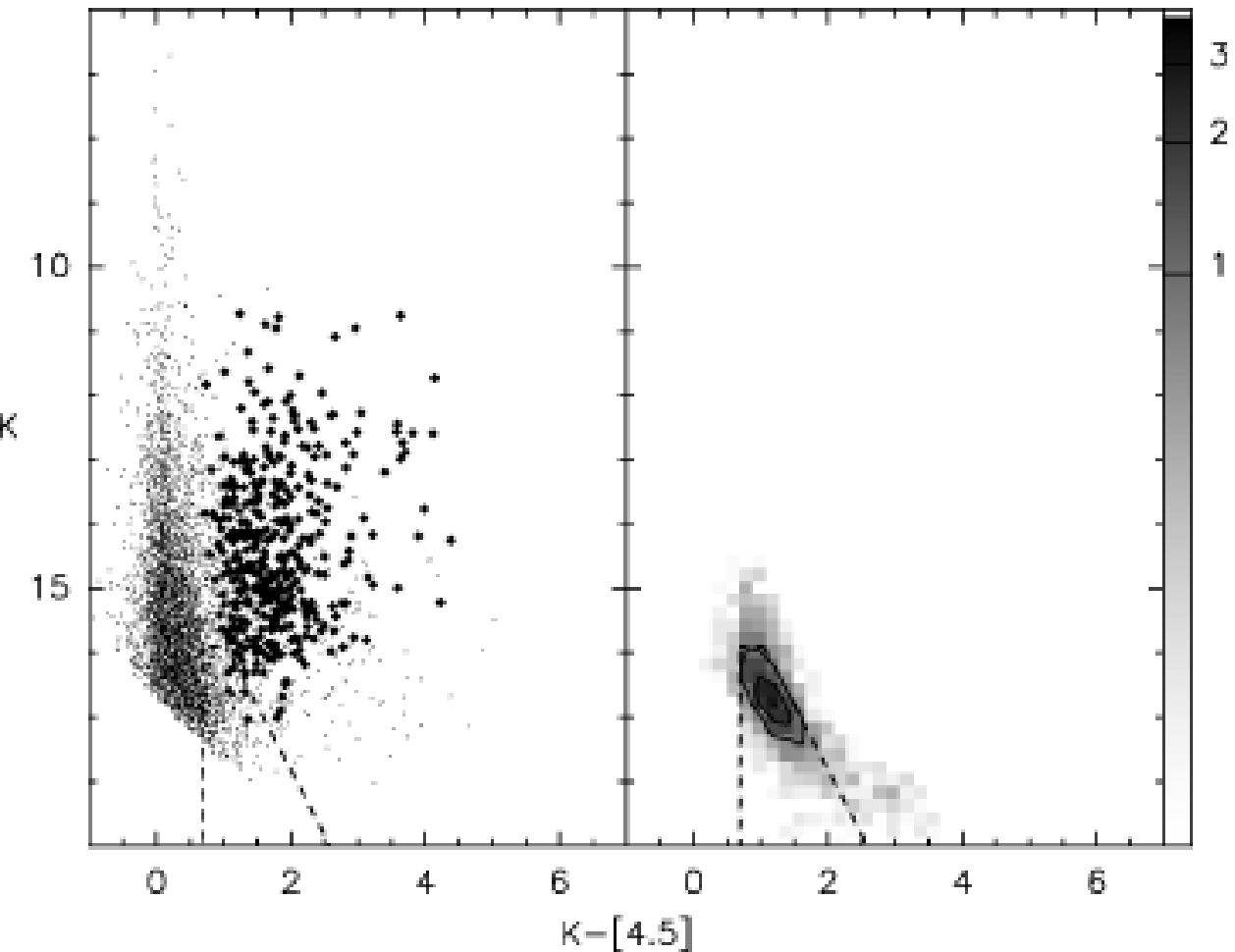}
\caption{\textbf{Upper-left}: Color-magnitude diagram for sources in \complex~detected in all IRAC bands. Circles show Class I sources, squares show Class II sources and points correspond to Class III sources. Dashed lines delimit background contaminated area. \textbf{Upper-right}: Distribution of galaxies from the IRAC Shallow Survey for the same color-magnitude diagram. Contours begin at 1.0 and increase by 1.0. \textbf{Lower-left}: Color-magnitude diagram for sources with H, K, and IRAC 4.5 $\mu$m detections. Sources with IR-excess are shown in diamonds. Dashed lines delimit background contaminated area. \textbf{Lower-right}: Distribution of Shallow Survey galaxies for the color-magnitude diagram. Contours begin at 1.0 and increase by 1.0.\label{background}}
\end{figure}

We used two different color-magnitude diagrams in our analysis: [4.5]-[8.0] vs. [4.5] for sources detected at least in the four IRAC bands, and K-[4.5] vs. K for the rest of the sources detected in K band and 4.5 $\mu$m. In order to compare our data with the Shallow sample, we normalized the number of galaxies by the ratio of the observed areas, and we added to the galaxy fluxes an extinction of A$_K=0.4$, corresponding to the median extinction in the \complex~complex. Figure~\ref{background} shows the diagnostics used to separate galaxies from YSO for sources with detections in the four IRAC bands (upper) and in the NIR+IRAC (lower). The left part of the diagram shows the sources detected in the \complex~complex, and the right part shows the distribution of galaxies from the Shallow survey normalized by the area. We defined the ``contaminated area'' as the area with more than 1 galaxy per color-magnitude bin. In order to minimize the extragalactic background, we exclude from our analysis all the sources detected inside the contaminated area in the color-magnitude diagram.

We estimate that before correction, the number of galaxies in our mosaics is around 30. After applying these background elimination criteria, we estimate that there are at most 3 galaxies left in our sample.

\subsection{Spatial Distributions of YSO}\label{section_distribution}
\subsubsection{Class I and Class II sources}\label{section_distribution_classes}
\begin{figure}
\epsscale{0.8}
\plotone{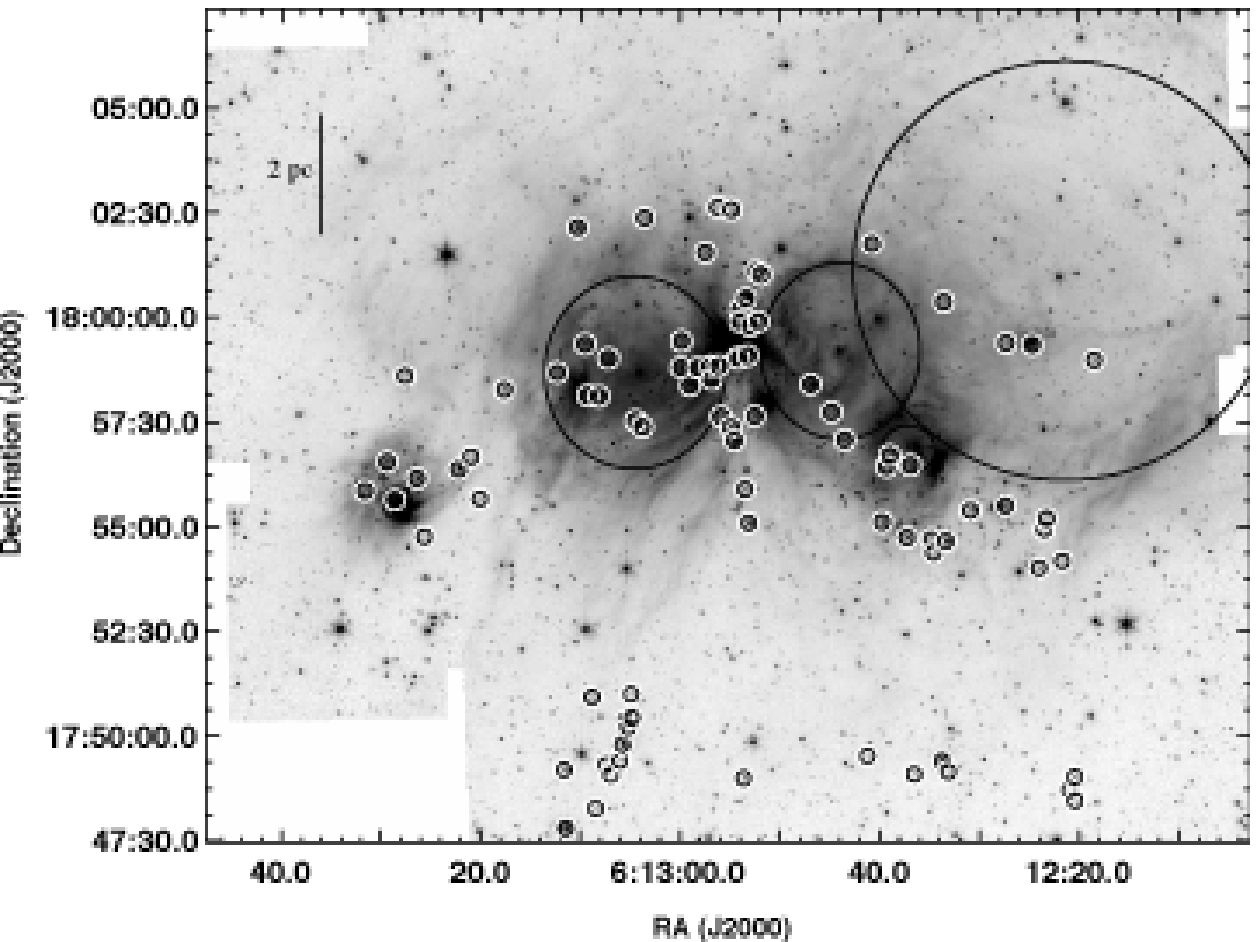}
\plotone{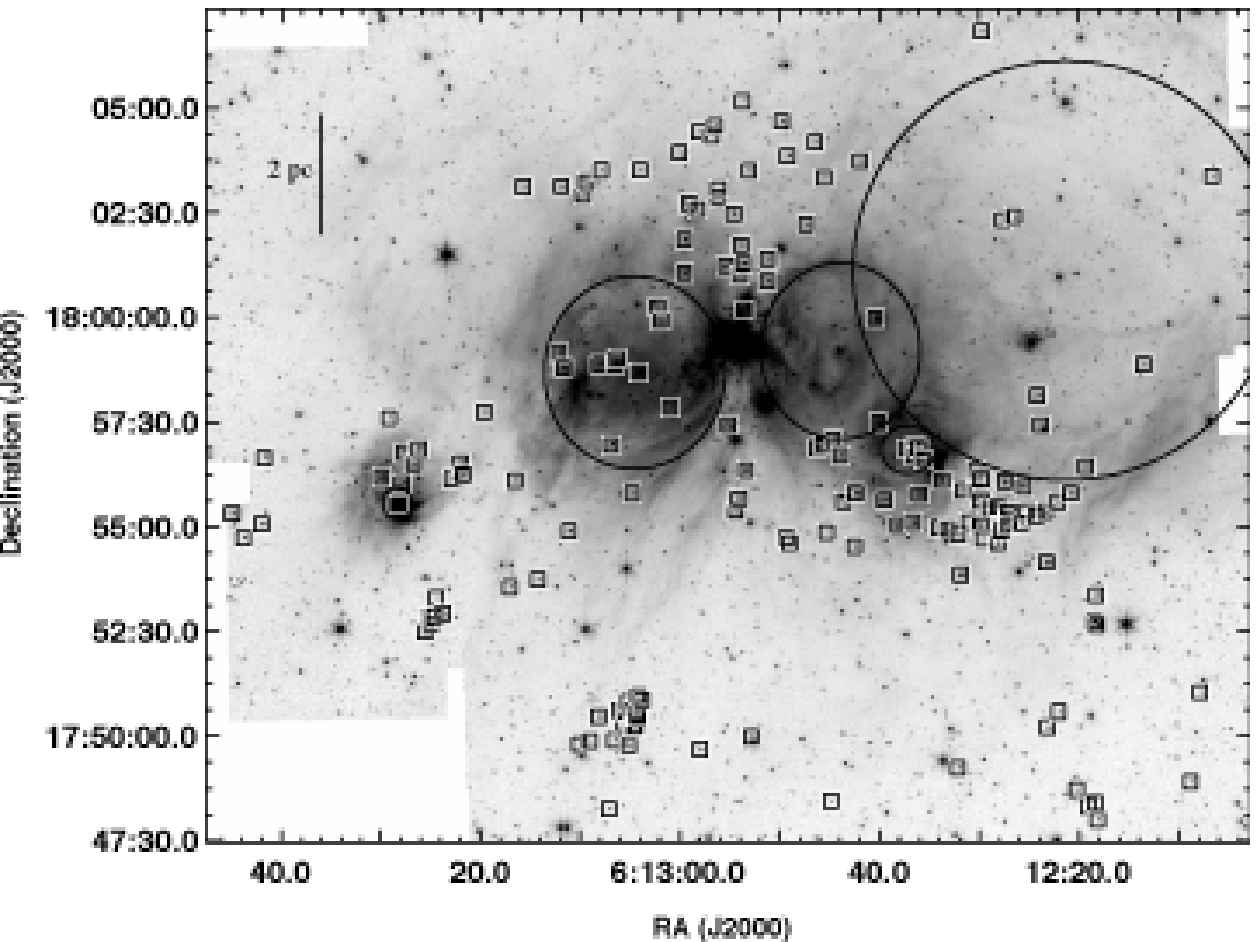}
\caption{\textbf{Upper:} Sources identified as Class I are shown marked with circles on the IRAC 4.5 $\mu$m mosaic of \complex. \textbf{Lower:} Sources identified as Class II (squares) on \complex. Large black circles correspond to the \hii~regions in the complex (from \citet{miz82})\label{irac_classes}}
\end{figure}

Figure~\ref{irac_classes} shows the spatial distribution of Class I and Class II sources, based on our classification in \S~\ref{section_classification}. In general, Class II sources are more dispersed than Class I. There is a clear concentration of Class I sources around S255-2 that is extended in the north-south direction, forming a filament of YSO between S255 and S257. Another, more dispersed group of Class I is located in S256 and to the south of S256 and S254. We also detected Class I sources in S258, in the cluster near the center of S254, and a filamentary group in the lower part of the mosaic. We do not detect Class II sources in the core of S255-2, but this may be due to the bright PAH emission in the area (see end of \S~\ref{7band}). We detected a high concentration of Class II sources in S256 and to the south between S254 and S256. There is also a halo of Class~II sources north from S255-2, spreading to the north between S255 and S257. There are more Class II sources towards S258, a small group located south from S258, and another clustered group in the lower part of the mosaic.

Since Class II sources are assumed to be more evolved than Class I, we can use the ratio of Class II to Class I sources as a proxy for the relative ages of the clusters. In this scheme, even though we are missing some Class II sources, the high concentration of Class I suggests that the youngest population is located between the \hii~regions S255 and S257 (see Table~\ref{clusters}).

\subsubsection{IR-excess sources} \label{section_density}
Figure~\ref{clusters_size} shows the spatial distribution of all the sources with IR-excess (from \S~\ref{IRexcess}). We detect a large number of YSO between S255 and S257 that were not detected previously due to the extended emission at 5.8 and 8.0 $\mu$m. The filament of YSO located between S255 and S257 is mixed to the south with another group of YSO concentrated in S256 and extended east-west to the south of S257 and S254. We also detected a large concentration of YSO in S258, and other four groups: one located south from S258, another south-east from the first one, another group inside S254, and a group in the south-west corner of the image.

\subsubsection{IR-excess sources surface density map} 
\begin{figure}
\epsscale{0.8}
\plotone{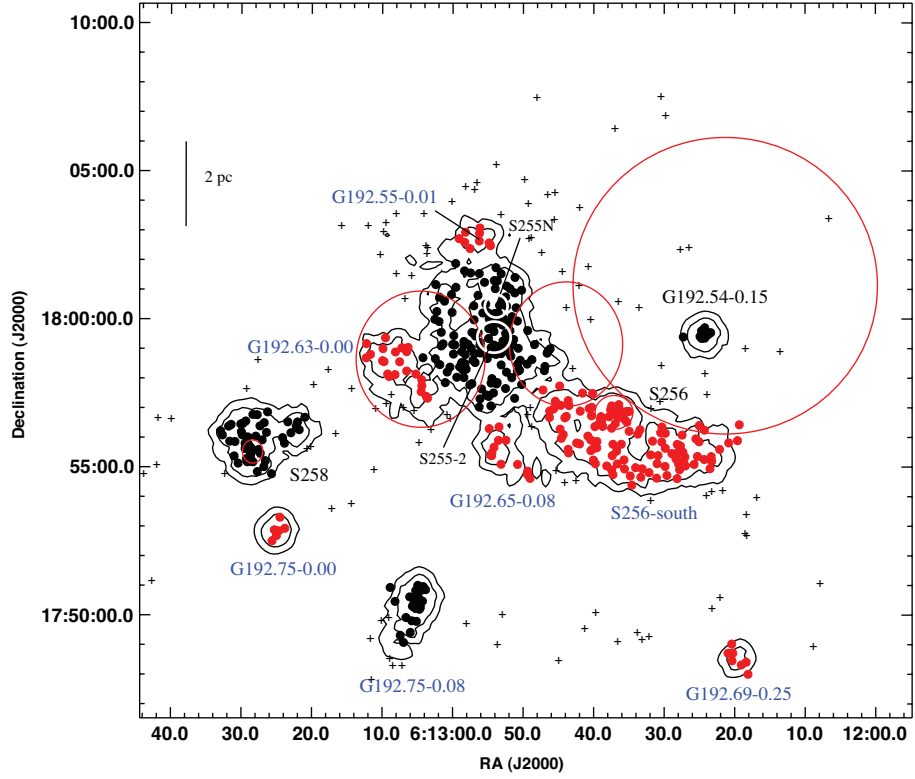}
\caption{Sources with IR-excess in the \complex~complex. Cluster members are shown in black and red points, the difference in color is just to separate the clusters. Isolated sources are shown in black plusses. Black contours represent the $n=5$ nearest neighbor surface density of YSO. Contour levels are at 5 and 10~stars~pc$^{-2}$. \hii~regions are shown in red color. New clusters are labeled in blue. White ovals in the center enclose clusters S255-2 and S255N.\label{clusters_size}}
\end{figure}

Having identified the young population of stars via their infrared excess, we studied the spatial distribution of YSO in the complex. Contours in Figure~\ref{clusters_size} correspond to the surface density map of sources with IR-excess. The map was created using a 3 arc-second grid. At each point of the grid, we calculated the surface density of stars given by:
\be
\sigma=\frac{N}{\pi r_{N}^2},
\ee
where $r_{N}$ is the distance to the $N=5$ nearest neighbor (NN). The contours identify clusters S255-2, S256, S258 and the cluster inside S254, but also show other concentrations of YSO in the complex towards the south of the mosaic. 

\begin{figure}
\epsscale{0.8}
\plotone{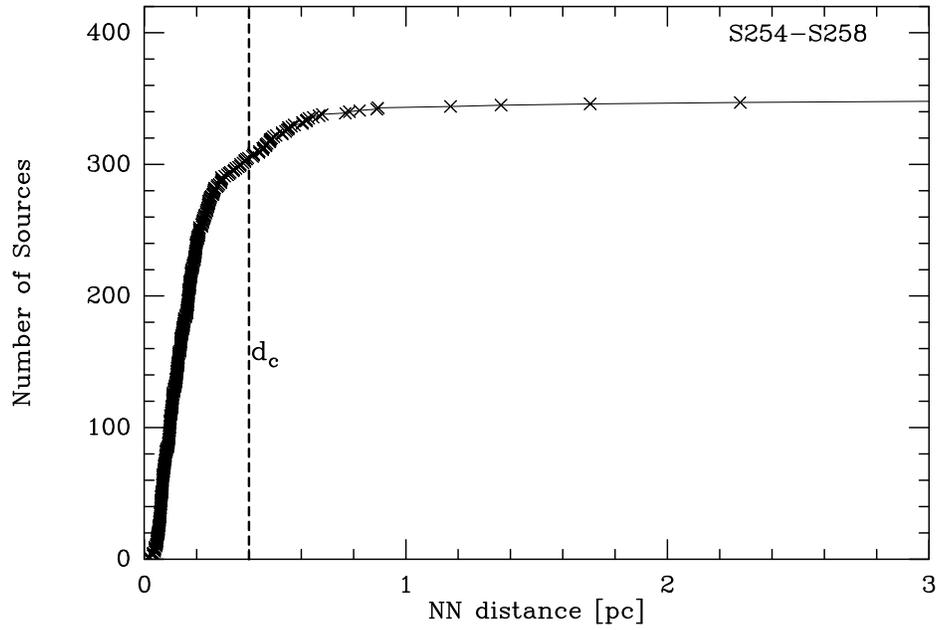}
\caption{Cumulative distribution of nearest-neighbor (NN) distances for sources with IR-excess in the \complex~complex. The low-density component starts at a distance of 0.01 degrees (or 0.4 pc at 2.4 kpc of distance). \label{dcritic}}
\end{figure}

Another feature seen in Figure~\ref{clusters_size} is that YSO clusters seems to be surrounded by a low-density distributed population of YSO. In order to determine the percentage of YSO located in clusters, we need to separate them from the low-density component. We establish the separation using the near neighbor cumulative distribution of sources with IR-excess. Figure~\ref{dcritic} shows the NN cumulative distribution for sources with IR-excess in the complex \complex. We have chosen (somewhat arbitrarily) the inflection point at 0.01 degrees (or 0.4 pc at 2.4 kpc of distance) as $d_c$, the maximum separation between cluster members.

Using $d_c$, and imposing a minimum number of members of 5 stars per cluster and a surface density of at least 5 stars per pc$^{2}$, we identified six new clusters of YSO in the complex and a group located south from S256 that may be part of the cluster in S256 (see Fig.~\ref{clusters_size}). The first new cluster (G192.75-0.00), is located south of S258. The second one, called G192.75-0.08, is located south-west from G192.75-0.00. The third (G192.63-0.00) is located east from S255-2, is more dispersed than the other two and includes the ionizing source of the \hii~region S255. The fourth cluster called G192.65-0.25, is located south-west from S256. The fifth and sixth clusters (G192.65-0.08 and G192.55-0.01) are located between the \hii~regions S255 and S257, south from S255-2 and north from S255N respectively. The number of members in each cluster, as well as the number of Class I and Class II sources per cluster are shown in Table~\ref{clusters}. We also identified the already known clusters S255-2, S256, S258, and G192.54-00.15, located inside \hii~region S254 \citep{dut01}. Finally, we calculated that the ratio of young stars in clusters vs. the total number of YSO is 0.8. If we change the value of $d_c$ in $\pm$20\%, the ratio change to 0.83 and 0.75 respectively.

\subsection{Cloud structure}

\subsubsection{K-band extinction map}
\begin{figure}
\epsscale{0.7}
\plotone{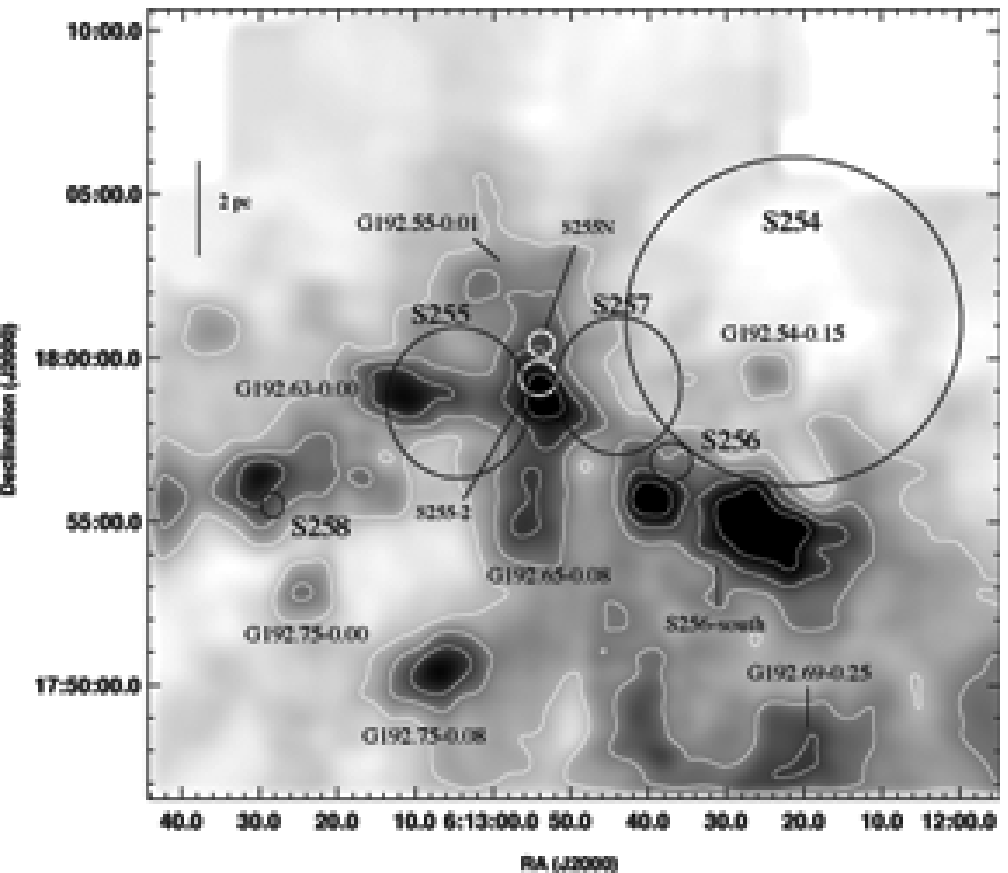}
\epsscale{0.4}
\plotone{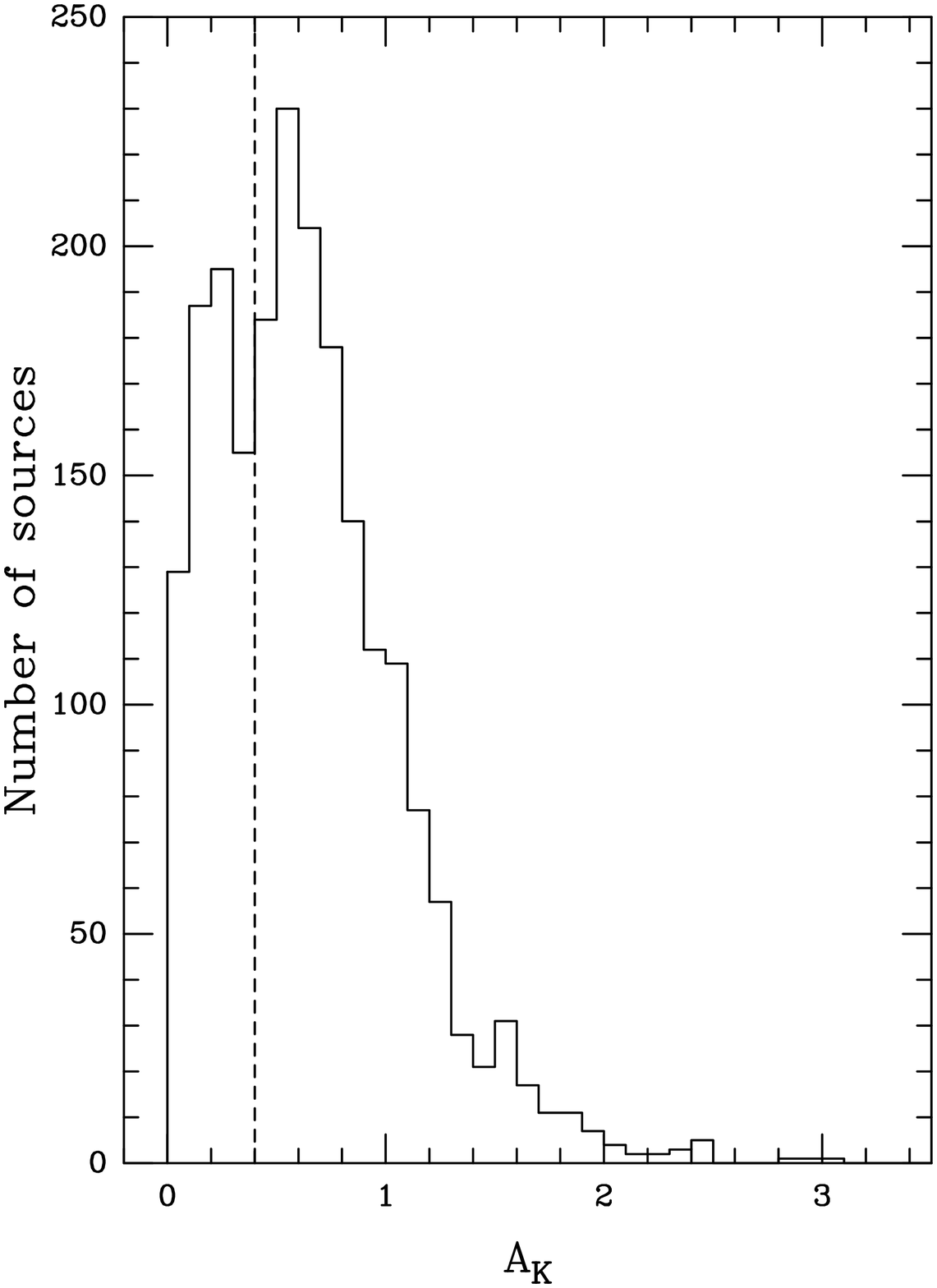}
\caption{\textbf{Upper}: K-band extinction map in the \complex~complex. Contours begin at 0.4 A$_K$ and increase by 0.2. Labels are the same as in Figure~\ref{clusters_size}. \textbf{Lower}: Distribution of A$_K$ for sources in the complex. Dashed line indicates the foreground extinction value adopted. \label{extinction}}
\end{figure}

The K-band extinction map of the complex is shown in Figure~\ref{extinction}. The map was created using the mean value of the K-band extinction of the nearest $N=10$ stars over a uniform 3 arc-seconds grid. The A$_K$ values were estimated by dereddening the sources with no IR-excess in the H-K vs. J-H color-color diagram, using the reddening law by \citet{ind05}. Finally, we convolved the resulting map with a 30 arc-seconds Gaussian kernel. The total number of sources used to build the map was 1937.

Since the complex is located at 2.4 kpc, foreground field stars will be reddened by interstellar extinction. The histogram in Figure~\ref{extinction} shows the distribution of A$_K$ in the complex. If there were no cloud in the field, the value of A$_K$ would rise, and then fall when it reaches our completeness limit, having a single peak. The presence of the molecular cloud splits the distribution of A$_K$ in two groups since the value of the extinction rises where the cloud is located, separating the foreground from the background stars. This separation is seen in the histogram at A$_K \sim 0.4$. This value is in good agreement with values of interstellar extinction at 2.4 kpc (0.15 mag per kpc in K-band, \citet{ind05}). Therefore, to eliminate the foreground contribution to the extinction, we only used stars with A$_K > 0.4$ in our extinction map. Also, since we are interested in the extinction due to only the molecular cloud (and not due to the ISM), we shifted A$_K$ down by 0.4 magnitudes. 

The A$_K$ values derived may be under-estimated due to the small number of background stars detected. The 90\% completeness limit in the K-band for our dataset is 16 magnitudes, corresponding to a K0 type star at 2.4~kpc from the Sun. For example, if we want to achieve an extinction of 3 magnitudes in the K-band, we have to detect a star of spectral type A0 at 2.4 kpc, or B8 at 3.0 kpc. In this mass range, the number of background stars available to measure high extinctions decreases rapidly.

\begin{figure}
\epsscale{0.8}
\plotone{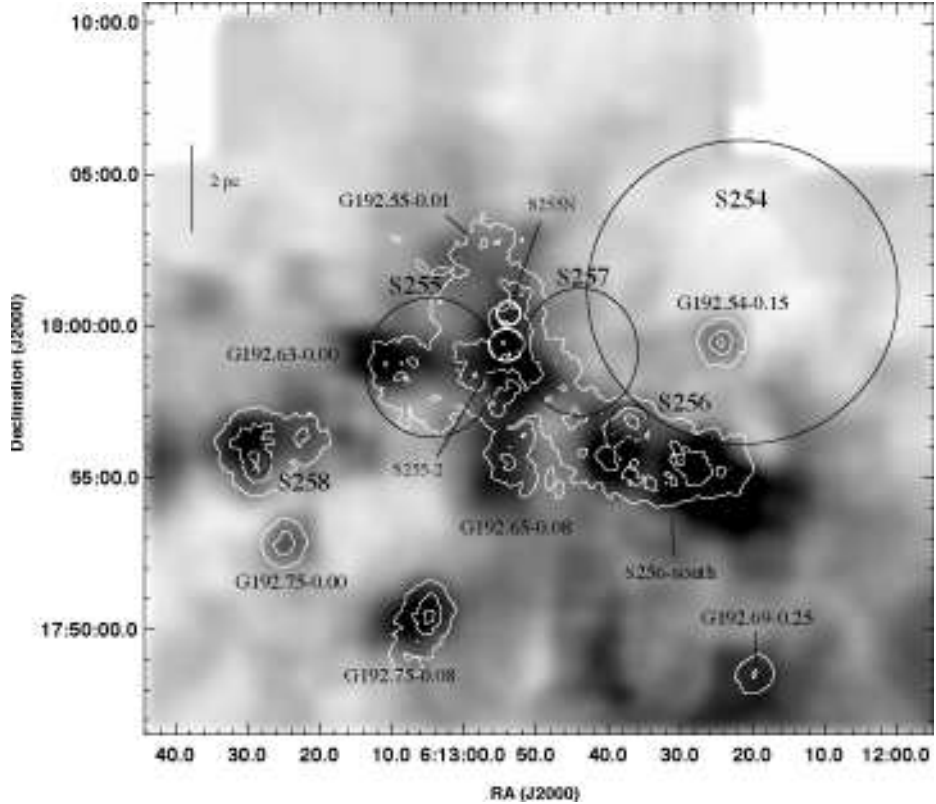}
\caption{K-band extinction map in the \complex~complex, overlaid with contours of the surface density of sources with IR-excess. Contour levels are at 5, 20 and 80~stars~pc$^{-2}$. Labels are the same as in Figure~\ref{clusters_size}. Note the $\sim 1$ arc-minute offset between the extinction and stellar density peaks in S256 and S258.\label{extinction_density}}
\end{figure}

The contours shown in Figure~\ref{extinction} start at A$_K$ of 0.4 and increase by increments of 0.2. The extinction map traces what appear to be three dusty filaments in the complex; extending north-south and centered on S255-2; south of S254 and S256, extending east-west; and from S255-2 to the east, in the middle of the \hii~region S255, where G192.63-0.00 is located. There are also regions with high extinction towards S258, G192.54-0.15, G192.75-0.00, and G192.75-0.08, and south from G192.69-0.25. Although the extinction map resembles the YSO density map presented in Figure~\ref{clusters_size}, we notice one difference. The peaks in the extinction and in the YSO density maps for S256 and S258 are separated by approximately 1 arc-minute (0.6 pc at a distance of 2.4 kpc), while for the other groups of stars the peaks seem to match in position (see Figure~\ref{extinction_density}). This difference could be explained by the fact that the \hii~regions are removing the dust from the clusters (see comparison with $^{13}\mathrm{CO}$ column density in next section).

\subsubsection{Molecular Gas}\label{section_molecular_gas}
We used the column density of $^{13}\mathrm{CO}$ and the velocity maps of $^{12}\mathrm{CO}$ to study the structure of the molecular cloud. We derived the $^{13}\mathrm{CO}$ column density from: 
\be
N(^{13}\mathrm{CO}) = 2.42\times10^{14}\left(\frac{\Delta v T_{ex}\tau}{1 - e^{-5.29/T_{ex}}}\right) cm^{-2},
\ee
where
\be
T_{ex} = \frac{5.53}{ln\left(1 + \frac{5.53}{T_A + 0.82}\right)},
\ee
\be
\tau = -ln\left(1 - \frac{T_{ex}}{\frac{5.29}{e^{5.29/T_{ex}}-1}-1}\right),
\ee
$T_A$ is the observed peak temperature of the $^{12}\mathrm{CO}$, and $\Delta v$ correspond to the full width at half maximum (FWHM) of the $^{13}\mathrm{CO}$ line \citep{hey89}. We obtained the FWHM by fitting a Gaussian profile to the velocity spectrum in each pixel\footnote{Based on code by R. Franco-Hernandez}. 

\begin{figure}
\epsscale{0.7}
\plotone{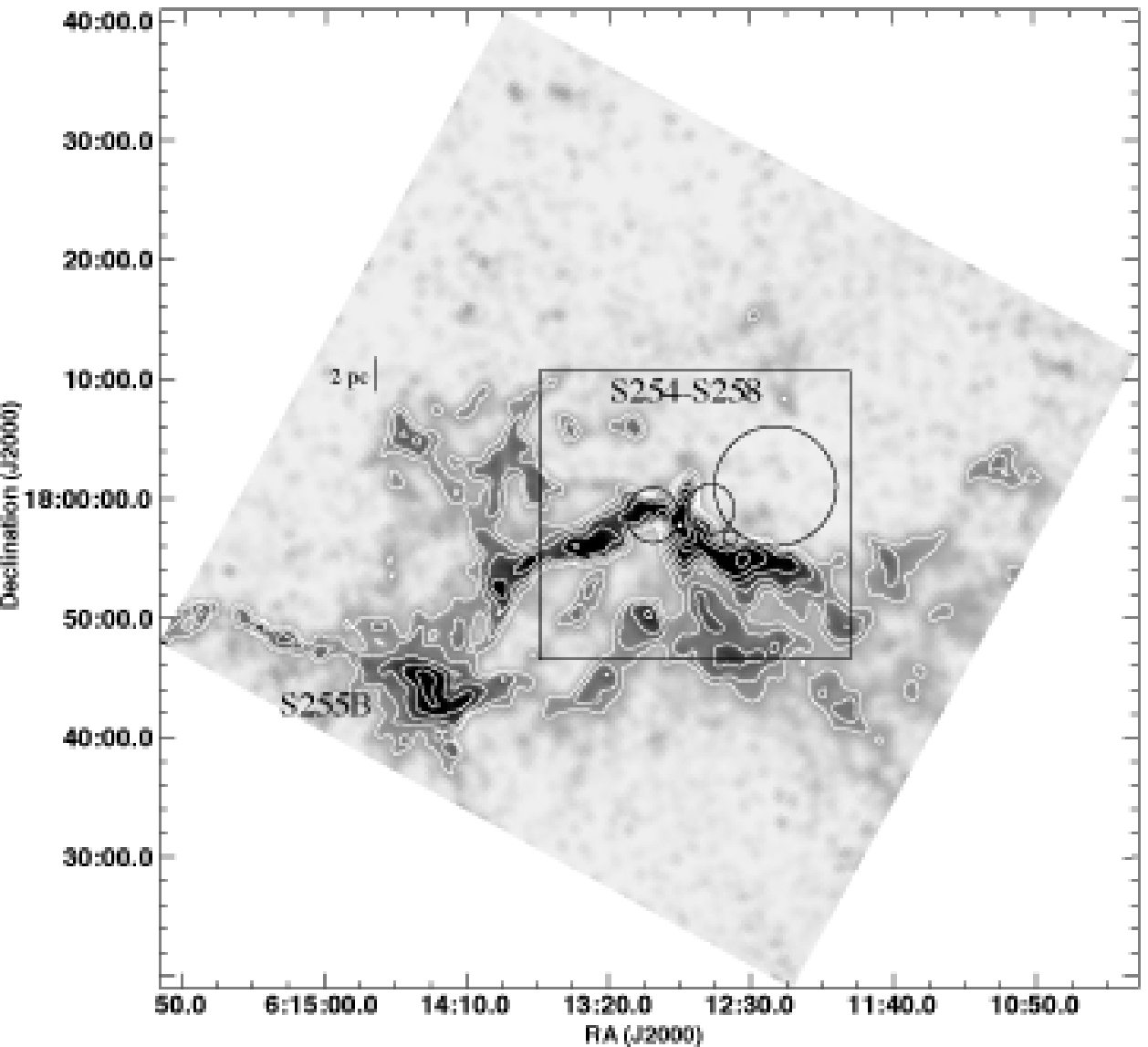}
\plotone{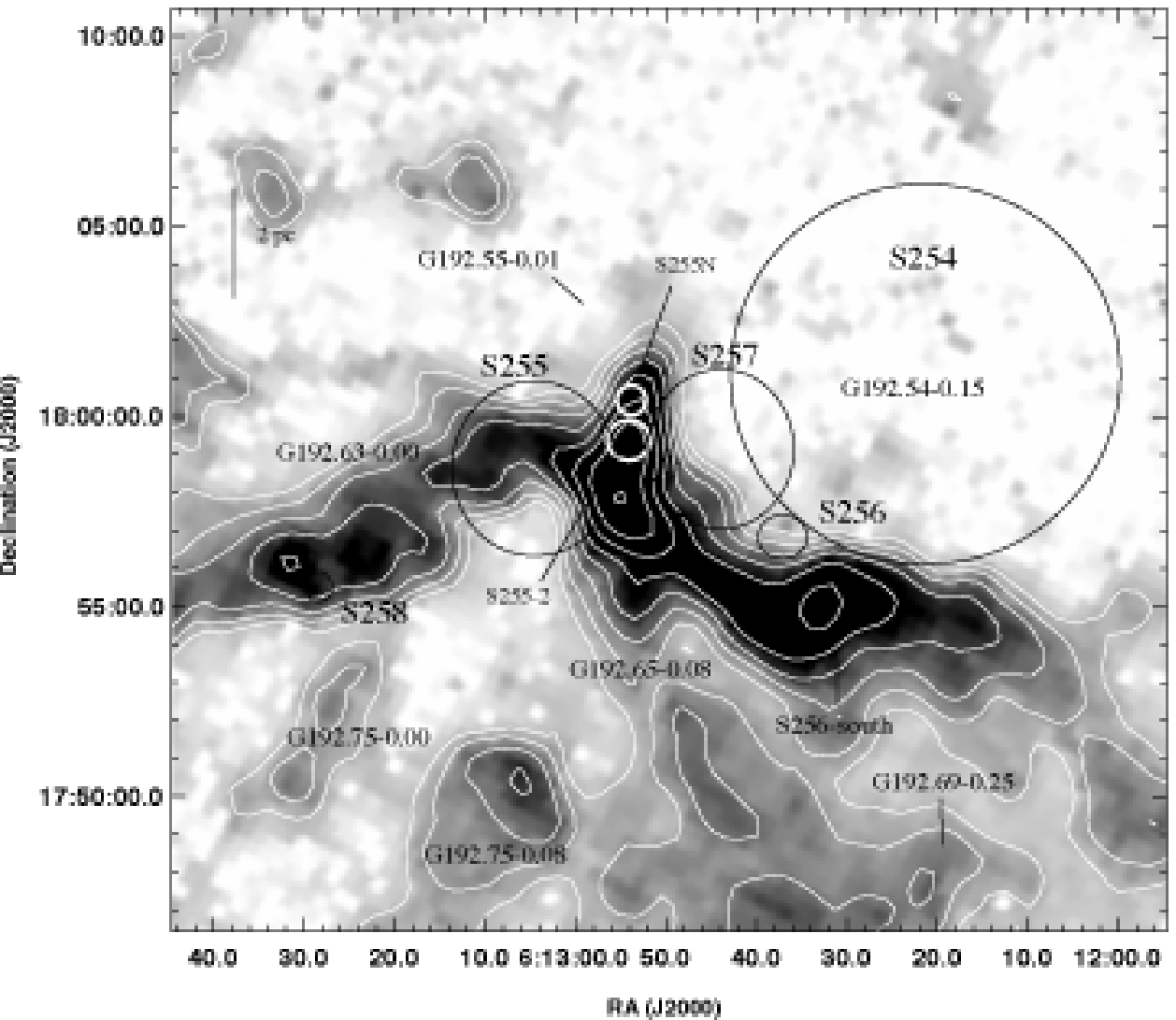}
\caption{\textbf{Upper}: Map of $^{13}\mathrm{CO}$ column density derived from the ratio of $^{12}\mathrm{CO}$ and $^{13}\mathrm{CO}$ intensities. Contours are spaced logarithmically between $6\times 10^{15}$ and $6\times 10^{16}$cm$^{-2}$. \textbf{Lower}: Close up of the boxed region having the same size as the K-band extinction map (shown in Figure~\ref{extinction}). Labels are the same as in Figure~\ref{clusters_size}. \label{13coden}}
\end{figure}

The $^{13}\mathrm{CO}$ column density map is shown in Figure~\ref{13coden}. There is a long (20pc) horizontal filament with a vertical component between S255 and S257. This filament is connected on the east side with the region S255B to the south \citep{bic03}, and with another vertical filament to the north. There is also more molecular material south from the \complex~complex. The expanded view in Figure~\ref{13coden} shows the column density in the same area as the K-band extinction map, \hii~regions and clusters are indicated in this figure. The column density peak is located between \hii~regions S255 and S257, forming a vertical filament in the middle of the \hii~regions. We detected molecular material in the locations of all the clusters in the complex. Compared with the map from \citet{hey89}, the column density values obtained are similar. In our case, we do detect molecular material in the location of cluster G192.75-0.00. The same for cluster G192.54-0.15, but with a lower column density. 

In general the $^{13}\mathrm{CO}$ density map agrees well with the K-band extinction map, showing the presence of gas around the locations where star formation is occurring, as well as in places without clusters but that are part of the molecular cloud, like south of S256 and west of G192.75-0.08. The column density peaks agree reasonably well with the extinction peaks for most of the clusters. In S255-2, the column density peak is located 1 arc-minute south of the extinction peak, this difference may be due a sensitivity problem in our extinction map caused by the non detection of background stars in places of high extinction.

\begin{figure}								
\begin{center}
\epsscale{1.0}
\plotone{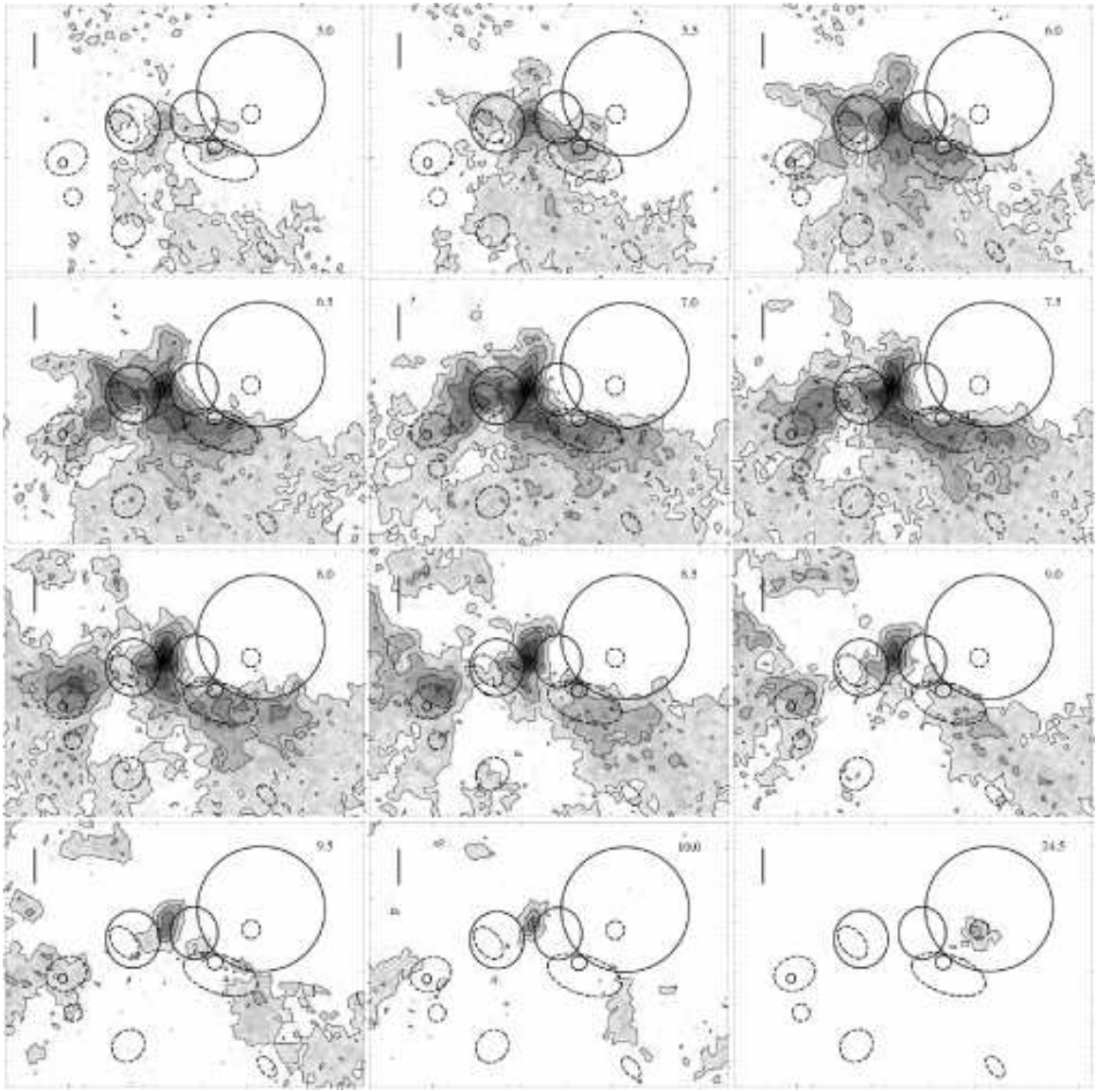}			
\caption{$^{12}\mathrm{CO}$ channel maps. The velocity (km~s$^{-1}$) is indicated by the number in the top-right corner. Contour levels begin at 3K, and increase by 3K. H$\scriptsize{\mathrm{II}}$ regions are indicated by black circles, the approximate size of the clusters is indicated by segmented ovals. Black line represents a scale bar of 2 pc. As shown also by \citet{hey89}, the bulk of the cloud is located at velocities between 5 and 10~km~s$^{-1}$. The only emission that matches the position of G192.54-0.15 is detected at 24.5~km~s$^{-1}$ (lower right).\label{12co}}
\end{center}
\end{figure}

In both the column density map as well as in the extinction map, there is no material inside the \hii~region S254 other than in the location of G192.54-0.15, suggesting that the \hii~region already removed most of the molecular gas and dust. If this is the case, the molecular material in G192.54-0.15 must likely show some evidence of interaction with the ionizing star (like a cometary structure) due to their proximity. Such an interaction is not seen in the IRAC mosaics, suggesting that the cluster may not be related with the complex. The $^{12}\mathrm{CO}$ velocity maps in Figure~\ref{12co} show that the main body of the molecular cloud has a V$_{\mathrm{LSR}}$ between 5 and 9~km~s$^{-1}$, while the $^{12}\mathrm{CO}$ emission in G192.54-0.15 has a V$_{\mathrm{LSR}}$ of 24-25~km~s$^{-1}$. Since it seems unlikely that the $^{12}\mathrm{CO}$ emission at that position is not related with G192.54-00.15, the difference in the velocity with respect to the main body of the molecular cloud suggests that the cluster is located in the background of the complex (see \S~\ref{section_little_cluster}).

Finally, we used the $^{13}\mathrm{CO}$ column density to estimate the gas mass for clusters in the complex. In order to do this, we enclosed the cluster members by an ellipsoid and then we calculated the mass inside the ellipsoid assuming an abundance of $^{12}\mathrm{CO}$ to $^{13}\mathrm{CO}$ of 50 \citep{isr03}, and of $^{12}\mathrm{CO}$ to H$_2$ of 8.5$\times 10^{-5}$ \citep{fre82}. Gas masses, and other properties of the clusters are shown in Table~\ref{clusters} and are discussed in \S~\ref{section_discussion}.

\section{Discussion}\label{section_discussion}

\subsection{Clusters in the complex}
We detected in total 510 sources with IR excess in the complex \complex, including 87 Class I, and 165 Class II sources. Of those, 398 are located in clusters (as described in \S~\ref{section_density}). Here, we discuss the global properties of the clusters.

\subsubsection{S255-2 \& S255N}
These are located between \hii~regions S255 and S257. We detected 23 Class I, 14 Class II, and 140 sources with IR-excess. We are likely to be missing some YSO due to the bright PAH emission (see \S~\ref{7band}). The core of S255-2 is saturated in the IRAC bands. The peak surface density of sources with IR-excess is 156 stars pc$^{-2}$, and the K-band extinction peak is 1.4 magnitudes, located 1 arc-minute south of S255-2. The peaks are coincident in location within 1 arc-minute (0.6 pc at a distance of 2.4 kpc). The $^{12}\mathrm{CO}$ spectrum peaks at 7.5~km~s$^{-1}$ for S255-2 and at 8.4~km~s$^{-1}$ for S255N. 

The detection of ultra-compact \hii~regions (UC\hii), outflows, and the high ratio of Class I to Class II suggest that S255-2 and S255N are clusters in an early stage of formation (see also \citet{cyg07}).

\subsubsection{S256 \& S256 south}
The YSO in S256 are more concentrated compared to S256-south. We detect 14 Class I, 44 Class II, and 115 sources with IR-excess. The surface density peak is 256~stars~pc$^{-2}$ and the peak in $A_K$ is 1.3 magnitudes, located 1 arc-minute south-west from the surface density peak. The $^{12}\mathrm{CO}$ spectrum shows a double peak in S256, at 6.1 and 8.5~km~s$^{-1}$, and a small component at 3.1~km~s$^{-1}$. For S256-south there is a single peak at 7.4~km~s$^{-1}$, and a small peak at 3.3~km~s$^{-1}$. The spectral type of the central source in the \hii~region S256 derived from our data is B0.9V. The source is classified as a Class~II.

\begin{figure}
\epsscale{0.8}
\plotone{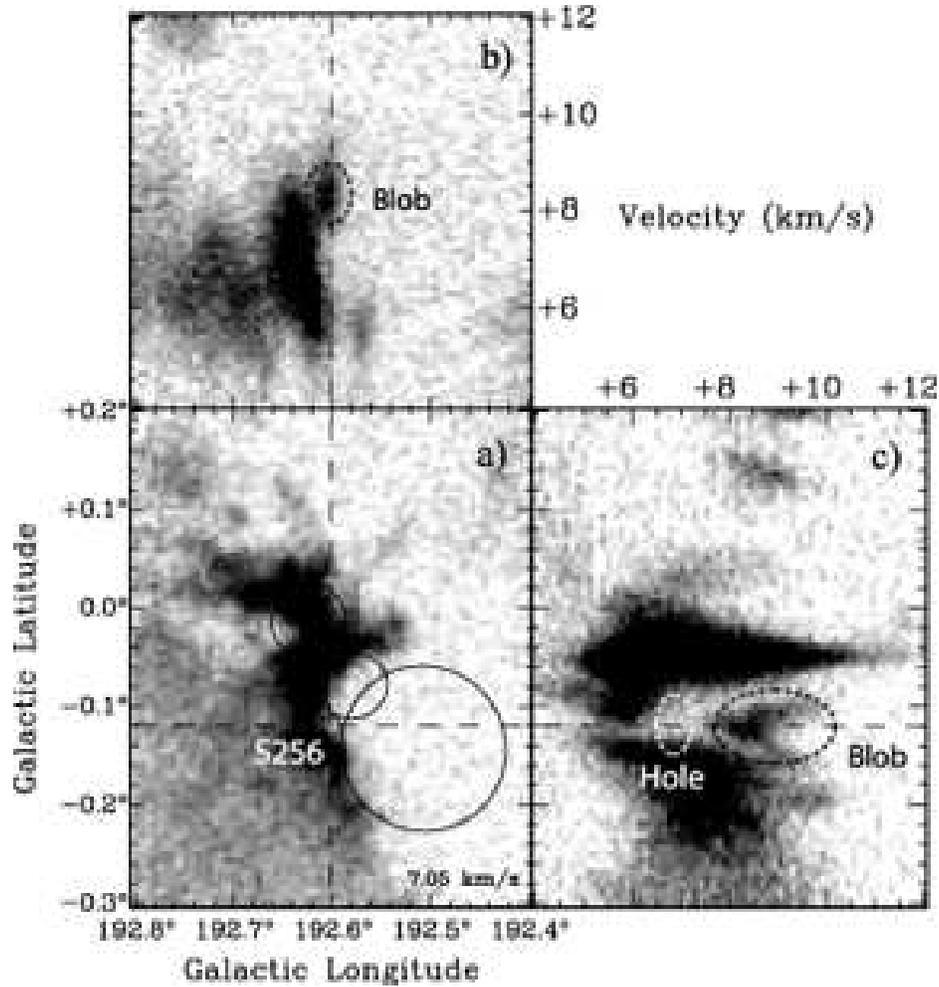}
\caption{$^{12}\mathrm{CO}$ position and position-velocity maps in galactic coordinates of the \complex~complex. In a), we show the emission map at 7 km s$^{-1}$. The black ovals represent the size of the H$\scriptsize{\mathrm{II}}$ regions. b) and c) are position-velocity maps in latitude and longitude respectively across the dashed lines in a). c) shows a hole in the molecular cloud located at the same position as S256, and a blob of gas along the same line of sight at 9 km s$^{-1}$.\label{s256}}
\end{figure}

Figure~\ref{s256} shows $^{12}\mathrm{CO}$ position-velocity maps of the area around S256. The peak at 8.5~km~s$^{-1}$ corresponds to a blob of molecular material that seems to have been ejected from the cloud. The mass of the blob is 30 M$_{\odot}$. If we assume a velocity of 2~km~s$^{-1}$ with respect to the main body of the cloud, the estimated kinetic energy is 1$\times10^{45}$~ergs (the energy radiated by an B0.9V star in 1$\times10^5$~yr is about 1$\times10^{46}$~ergs).

The presence of an \hii~region instead of an UC\hii~region, and also the ratio of Class I to Class II sources suggest that S256 and S256-south are more evolved than S255-2. 

\subsubsection{S258}
We detect 7 Class I, 9 Class II, and 49 sources with IR-excess. The peak surface density is 298~stars~pc$^{-2}$ and the peak $A_K$ is 1.0 magnitudes. The $^{12}\mathrm{CO}$ spectrum has a strong peak at 8.0~km~s$^{-1}$, and a small peak 3.8~km~s$^{-1}$. The spectral type of the central source in the \hii~region derived from our data is B1.5, corresponding to a Class~II. 

Similar to cluster S256, S258 seems more evolved than S255-2. The location of the central source in the \hii~region indicates that it is likely the ionizing source. The $^{13}\mathrm{CO}$ column density map shows that S258 is part of the same filament where S255-2 and S256 are located.

\subsubsection{G192.54-0.15}\label{section_little_cluster}
This cluster is located inside the \hii~region S254. We detect 2 Class I, 0 Class II, and 9 sources with IR-excess. The peak in density is 166~stars~pc$^{-2}$ and the peak $A_K$ is 0.7 magnitudes. The $^{12}\mathrm{CO}$ spectrum peaks at 24.5~km~s$^{-1}$. Since the cluster is located near the center of the \hii~region S254, where most of the molecular material has been evacuated, we are able to detect it through the molecular cloud.

The kinematic distance for G192.54-0.15, using the rotational model from \citet{bra93} is about 9 kpc (the uncertainty in the estimation of the kinematic distance toward the anti-center of the Galaxy may be as much as a factor of 2, \citet{rei08}). At that distance, the cluster will be located in the outer parts of the Galaxy, having an IRAC luminosity 4 times less than the trapezium in Orion. There is evidence of molecular material in the same direction with similar V$_{\mathrm{LSR}}$ on a larger scale \citep{car95}. G192.54-0.15 may be part of star formation occurring in those molecular clouds, and it could be an important object for further study of star formation in the outer Galaxy.

\subsubsection{G192.75-0.00}
This cluster is located to the south of S258. We detect 0 Class I and 7 Class II sources. The peak in surface density is 76~stars~ pc$^{-2}$ and the peak in $A_K$ is 0.7 magnitudes. The $^{12}\mathrm{CO}$ spectrum peaks at 8.1~km~s$^{-1}$.

The absence of Class I objects suggests that this cluster may be older than S256 and S258. There is no signature of an \hii~region, implying that this cluster contains low mass stars.

\subsubsection{G192.75-0.08}
Located south-east of G192.75-0.00. We detect 25 sources with IR-excess, 9 Class I and 15 Class II. The peak in surface density is 238~stars~pc$^{-2}$ and the peak in $A_K$ is 1.1 magnitudes. The $^{12}\mathrm{CO}$ spectrum peaks at 7.2~km~s$^{-1}$.

As for G192.75-0.00, in G192.75-0.08 there is no signature of massive star formation. But the presence of Class~I sources suggests that it is younger than G192.75-0.00. The value of $A_K$, as well as the surface density of gas, is higher than in G192.75-0.00.

\subsubsection{G192.63-0.00}
This cluster is located inside \hii~region S255. We detect 7 Class I, 5 Class II and 22 sources with IR-excess. The peak in surface density is 39~stars~pc$^{-2}$ and the peak in $A_K$ is 1.0 magnitudes. The $^{12}\mathrm{CO}$ spectrum peaks at 6.9~km~s$^{-1}$. The ionizing star of S255 is located inside this cluster. The spectral type of the ionizing star derived from our data is B0.0, and is classified as a Class~II. 

The IRAC classification of the ionizing star in S255 (Class II) suggests that it is in an earlier stage than the ionizing star in S257 (Class III). In this context, we may assume that other YSO formed together with the ionizing star of S255 are also in an earlier stage than the YSO formed with the ionizing star of S257. In this case, we would expect to detect more sources with IR-excess in S255 than in S257. In fact, we detect almost 3 times more IR-excess sources inside \hii~region S255 than in S257. We hypothesize that G192.63-0.00 is the natal cluster of the ionizing star in S255.

\subsubsection{G192.69-0.25}
This cluster is located south-west from S256. We detect 7 sources with IR-excess; 3 Class I and 4 Class II. The peak in surface density is 27~stars~pc$^{-2}$ and the peak in $A_K$ is 0.8 magnitudes. The $^{12}\mathrm{CO}$ spectrum peaks at 7.0~km~s$^{-1}$.

Since this cluster is located at the edge of our mosaic, we may be missing some members. The $^{13}\mathrm{CO}$ column density map shows more molecular material in the location of the cluster south of our mosaic.

\subsubsection{G192.65-0.08 \& G192.55-0.01}
These two clusters are located between \hii~regions S255 and S257. G192.65-0.08 is located south from S255-2. We detect 13 sources with IR-excess; 2 Class I and 6 Class II. The peak in surface density is 37~stars~pc$^{-2}$ and the peak in $A_K$ is 0.9 magnitudes. The $^{12}\mathrm{CO}$ spectrum peaks at 7.0~km~s$^{-1}$. G192.55-0.01 is located north from S255N. We detect 9 sources with IR-excess; 2 Class I and 5 Class II. The peak in surface density is 30~stars~pc$^{-2}$ and the peak in $A_K$ is 0.7 magnitudes. The $^{12}\mathrm{CO}$ spectrum peaks at 6.3~km~s$^{-1}$.

\subsection{Star formation efficiency}\label{section_efficiency}
In order to estimate the star formation efficiency of clusters:
\be
\epsilon = \frac{M_{stars}}{M_{stars} + M_{gas}},
\ee
 we need to estimate the mass of gas and stars in the clusters. The mass of gas was calculated using the $^{13}\mathrm{CO}$ column density (see \S~\ref{section_molecular_gas}). To estimate the mass of stars in clusters we count the number of stars and assume each one has a mass of 0.5 M$_{\odot}$. To count the number of stars we must correct for two sources of incompleteness: 1) cluster members without IR-excess are not identified by our method, and 2) we do not detect cluster members fainter than 16 magnitudes in the K-band. To account for cluster members without IR-excess we calculated the K-band surface density in a region of the map without IR-excess sources (control field), then we calculated the number of cluster members by subtracting the mean surface density (normalized by the area) of the control field from the mean surface density of the clusters. To estimate the number of cluster members fainter than 16 magnitudes in K-band, we created a 1~Myr old synthetic cluster having an initial mass function (IMF) given by the Orion nebula cluster. We compared the observed K-band luminosity function (KLF) of a region including S255-2, S255N, S256 and S258 with the resulting KLF of synthetic clusters with ages of 0.5, 1, 3, and 5~Myr \citep{mue00}. We found that the 1~Myr old cluster best represents the young population in those regions. Then, using the synthetic 1~Myr old cluster, we estimated the percentage of cluster members brighter than our completeness limit (16 magnitudes in K-band), at a distance of 2.4~kpc and assuming an extinction of 0.4 magnitudes in the K-band. We found that this percentage is around 50\%, therefore the corrected number of cluster members is two times the number of observed members. We then estimated the mass of stars in clusters by multiplying the corrected number of cluster members by 0.5, which corresponds to the average mass of stars.

Star formation efficiencies for clusters in the complex are shown in the last column of Table~\ref{clusters}. We estimate an uncertainty of a factor of 2 in the values of the efficiency. The values are less than 20\% and comparable with those in other regions of star formation \citep{elm00} for most of the clusters with the exception of G192.54-0.15, G192.75-0.00, and G192.55-0.01. The cluster G192.54-0.15 has an efficiency of 54\% at 2.4~kpc from the Sun. The efficiency of this cluster at 9~kpc from the Sun is 1\%. 

\subsection{Ages of the \hii~regions and relative ages of clusters}
We considered the expansion time of the \hii~regions using the relation for the second expansion of the Str$\ddot{\textrm{o}}$mgren sphere given by \citet{sta04},
\be
t = \frac{4r_\mathrm{S}}{7c}\left[\left(\frac{r}{r_\mathrm{S}}\right)^{7/4} - 1\right], 
\ee
where the Str$\ddot{\textrm{o}}$mgren radius is:
\be
r_\mathrm{S} = \left(\frac{3\mathcal{N}}{4\pi\alpha n_H^2}\right)^{1/3},
\ee
$r$ is the size of the \hii~region, $c$ is the velocity of the sound, and $\alpha$ is the recombination coefficient \citep{peq91}. The sizes of the \hii~regions were estimated using H$\alpha$ imaging from \citet{miz82}. We assumed a Hydrogen density of $n_H=3\times 10^4$ cm$^{-3}$ \citep{wri81}, a gas temperature of $10^4$ K and a sound speed of 11~km~s$^{-1}$. We use the spectral type given by our data (listed in Table~\ref{ionizing_stars}) to estimate the amount of ionizing photons $\mathcal{N}$. The estimated ages for the \hii~regions in ascending order are: S258 (1$\times10^5$yr), S256 (2$\times10^5$yr), S255 (1.5$\times10^6$yr), S257 (1.6$\times10^6$yr), and S254 (5.0$\times10^6$yr). 

We used the ratio of Class I to Class II sources (I/II) to estimate the relative ages of clusters in the complex. If we assume that all the stars in a cluster were formed at approximately the same time, and since Class I objects are in an earlier stage of formation than Class II objects, we expect a high I/II ratio in young clusters. We also used the ratio of Class I to IR-excess sources (I/N$_{\textrm{\tiny{IR}}}$) to probe clusters with small number of members. The ratios I/II and I/N$_{\textrm{\tiny{IR}}}$ per cluster are shown in Table~\ref{clusters}. Even though the relative ages for the clusters in the complex given by both ratios is different, they give the same order from youngest to oldest in clusters S255-2, S258 and S256.

\subsection{Star formation in S254-S258: the overall picture}
We estimated the ages of the ionizing stars in the complex using the expansion time of their Str$\ddot{\textrm{o}}$mgren sphere. From young to old the \hii~regions are ordered as follows: S258, S256, S255, S257, and S254. Since the spectral type of the ionizing star in S255 and S257 is similar (the same for S256 and S258, see Table~\ref{ionizing_stars}), we can assume that both stars are evolving with similar time-scales. In this context, the age estimated using the Str$\ddot{\textrm{o}}$mgren sphere is consistent with their IRAC+NIR classification (Class~II for S255 and Class~III for S257). We found the same for the ionizing stars in S256 (Class~II) and S258 (Class~I). Also, the relative ages estimated for clusters S256 and S258 using the ratios I/II and I/N$_{\textrm{\tiny{IR}}}$ (S258 relatively younger than S256) is consistent with the ages estimated from the \hii~regions.

Both the column density map and the extinction map show that most of the gas and dust inside the \hii~region S254 has been removed. For \hii~regions S255 and S257, there is still some molecular material inside, while S256 and S258 are still embedded in the molecular cloud. This sequence is consistent with the evolution of the complex suggested by the ages of the \hii~regions. The column density map and the extinction map show that a filament of material is still present inside the \hii~region S255, while in S257 there is less molecular material. This is expected if S257 is older than S255 (since the spectral type of the ionizing stars is similar, the evolution of the \hii~regions should also be similar). This is consistent with the difference in the IRAC+NIR classification of the ionizing sources (S255 in an earlier stage than S257), and with the detection of 3 times more IR-excess sources inside S255 than in S257. Those sources may be part of the cluster where the ionizing star of S255 was born. 

The K-band extinction map shows that the extinction in clusters S256 and S258 peaks in a different location than the surface density of YSO. This is similar to Orion, where part of the star forming region has been exposed by the action of the massive stars. In the case of S255-2, the \hii~region is not formed yet, and for G192.75-0.00, G192.75-0.08, and G192.69-0.25, there is not evidence of massive stars in those clusters.

Using the $^{12}\mathrm{CO}$ velocity maps in Figure~\ref{12co} and the IRAC mosaics, we can study the spatial distribution of the \hii~regions in the complex. The molecular material accumulated between \hii~regions S255 and S257 suggests that those \hii~regions are located at the same distance from the Sun. For S254, the absence of accumulated material between S257 and S254 or any evidence of interaction with S257 or evidence that star formation happened between S254 and S257 (like in between S255 and S257), suggests that it is located outside the plane of S255 and S257. Finally, the relative V$_{\mathrm{LSR}}$ peaks between G192.63-0.00, S255-2, and S255N (6.9, 7.5, and 8.4~km~s$^{-1}$ respectively) suggests that star formation is propagating toward the back of the \hii~regions S255 and S257, (the relative kinematic distance between those clusters is more than 150~pc).

\section{Conclusions}
We used Spitzer-IRAC and Flamingos-NIR data to identify YSO in the massive star forming complex \complex. We used the IRAC slope of the SED \ai~to classify YSO into Class I and Class II sources, and the K-4.5 $\mu$m v/s H-K color-color diagram to identify sources with IR-excess. We detected in total 87 Class I, and 165 Class II sources. Also, we identified in total 510 sources with IR-excess.

We used the surface density map of sources with IR-excess, and NN analysis to study the spatial distribution of YSO. We discovered six unknown clusters of YSO in the complex: G192.63-0.00, G192.75-0.00, G192.75-0.08, G192.69-0.25, and clusters G192.65-0.08 and G192.55-0.01 that are located between \hii~regions S257 and S255. We also discovered a southern component of the cluster S256. We found that 80\% of the IR-excess sources are located in clusters, while the rest is distributed in a more isolated component.

We estimated the ages of the ionizing stars in the complex using the expansion time of their Str$\ddot{\textrm{o}}$mgren sphere. The result is consistent with the IRAC+NIR classification of the ionizing stars, and with the relative ages estimated using the I/II and I/N$_{\textrm{\tiny{IR}}}$ ratios. The ages estimated for the \hii~region are: S258 (1$\times10^5$yr), S256 (2$\times10^5$yr), S255 (1.5$\times10^6$yr), S257 (1.6$\times10^6$yr), and S254 (5.0$\times10^6$yr).

We used the $^{13}\mathrm{CO}$ column density and the A$_K$ extinction maps to study the structure of the molecular cloud. The main body of the molecular cloud has V$_{\mathrm{LSR}}$ between 5-9~km~s$^{-1}$. The \hii~regions S255 and S257 are located in the same plane while S254 seems to be located behind S255 and S257. 

The V$_{\mathrm{LSR}}$ of cluster G192.54-0.15 suggests that it is located in the background of the complex. The double peak in the $^{12}\mathrm{CO}$ spectrum of S256 and $^{12}\mathrm{CO}$ position-velocity maps of S256 suggests that a blob of molecular material is being pushed to the back of the molecular cloud by the \hii~region.\\

\acknowledgments
This work is based in part on observations made with the Spitzer Space Telescope, which is operated by the Jet Propulsion Laboratory, Caltech, under a contract with NASA. Support for this work was provided by NASA through a contract issued by JPL/Caltech. We also thank NOAO for their student thesis support. We are grateful to Lynne Hillenbrand for facilitating the use of NIRC data in S255-2, August Muench for providing the synthetic clusters, Robert Gutermuth for helpful discussions on clustering, and Tracy Huard for helpful discussions on extinction maps. We are grateful to the staff at the MMT for assistance with the Hectospec observations: Perry Berlind, Mike Calkins, Mike Alegria, John McAfee, and Ali Milone. We thank the Telescope Data Center at the CfA for creating the Hectospec pipeline and archive, and Susan Tokarz for reducing the data. We thank Jesus Hernandez for providing the SpTclass pipeline to classify the Hectospec spectra. Xavier Koenig provided comments on the manuscript. The Five College Radio Astronomy Observatory was supported by NSF grant AST 0540852. Chris Brunt is supported by an RCUK Fellowship at the University of Exeter, UK.




Facilities: {\it Facilities:} \facility{Spitzer (IRAC)}, \facility{KPNO:2.1m (Flamingos)}, \facility{Keck:I (NIRC)}, \facility{FCRAO (SEQUOIA)}, \facility{MMT (Hectospec)}, \facility{IRTF (SpeX)}

\clearpage
\input{tab1.tex}
\clearpage
\input{tab2.tex}

\clearpage
\input{tab3.tex}

\clearpage
\input{tab4.tex}




\clearpage

\clearpage

\clearpage

\end{document}

%% file: tab1.tex
\begin{deluxetable}{lccccccccccccc} 
\tabletypesize{\footnotesize}
\tablecaption{Summary of detections\tablenotemark{a} and areas observed per filter\label{stars_detected}}
\tablewidth{0pt}
\tablehead{
\colhead{} & 
\colhead{J} & 
\colhead{H} & 
\colhead{K} & 
\colhead{H\tablenotemark{b}} & 
\colhead{K\tablenotemark{b}} & 
\colhead{JHK\tablenotemark{c}} & 
\colhead{[3.6]} & 
\colhead{[4.5]} & 
\colhead{[5.8]} & 
\colhead{[8.0]} & 
\colhead{All IRAC} & 
\colhead{HK[4.5]\tablenotemark{c}} 
}
\startdata
Detections        & 5941 & 5966 & 7430 & 152 & 145 & 4225 & 6131 & 6391 & 1472 & 853 & 462 & 3838 \\ 
Limiting magnitude\tablenotemark{d}&  16.0& 16.4 & 16.0 & 13.6 & 13.8&-&  14.0 &  13.5 & 11.0 & 10.0 &- &-\\
Area [arc-min$^2$]&  530 &  530 &  530 &   1 &   1 &  530 &  500 &  570 &  500 & 570 & 430 &  500 \\
\enddata
\tablenotetext{a}{Detections with error $<$ 0.2 magnitudes.}
\tablenotetext{b}{NIRC data.}
\tablenotetext{c}{Including Flamingos and NIRC data.}
\tablenotetext{d}{Magnitude for 90\% completeness.}
\end{deluxetable}

%% file: tab2.tex
\begin{deluxetable}{ccrrrrrrrrrrrrrrr} 
\tabletypesize{\footnotesize}
\tablecaption{Photometry\tablenotemark{a}\label{fluxes}}
\tablewidth{0pt}
\tablehead{
\colhead{RA (2000)} & 
\colhead{Dec (2000)} & 
\colhead{J} & 
\colhead{err} & 
\colhead{H} & 
\colhead{err} & 
\colhead{K} & 
\colhead{err} & 
\colhead{[3.6]} & 
\colhead{err} & 
\colhead{[4.5]} & 
\colhead{err} & 
\colhead{[5.8]} & 
\colhead{err} & 
\colhead{[8.0]} & 
\colhead{err} & 
\colhead{Class} 
}
\startdata
93.36916  & 17.92738 &  14.43 &  0.01 &  12.91 &  0.01 &  12.02 &  0.01 &  11.57 &  0.04 &  11.17 &  0.04 &  10.04 &  0.13 &  8.26 &  0.13 & I \\
93.28009  & 17.98375 &  17.91 &  0.08 &  15.78 &  0.03 &  14.16 &  0.01 &  12.09 &  0.02 &  10.94 &  0.02 &  9.76  &  0.02 &  8.59 &  0.03 & I \\
93.10282  & 17.98915 &  15.86 &  0.03 &  14.71 &  0.02 &  13.65 &  0.02 &  12.39 &  0.07 &  11.79 &  0.05 &  10.21 &  0.15 &  8.68 &  0.19 & I \\
93.14838  & 17.94358 &  15.72 &  0.01 &  13.90 &  0.01 &  12.77 &  0.01 &  11.12 &  0.02 &  10.35 &  0.01 &  9.85  &  0.04 &  9.01 &  0.09 & II \\
93.15528  & 17.94797 &  15.46 &  0.02 &  14.03 &  0.01 &  13.08 &  0.01 &  11.71 &  0.02 &  11.07 &  0.02 &  10.74 &  0.05 &  9.61 &  0.08 & II \\
93.15188  & 17.94479 &  16.17 &  0.04 &  14.22 &  0.01 &  13.04 &  0.01 &  11.85 &  0.02 &  11.29 &  0.02 &  10.81 &  0.05 &  9.87 &  0.11 & II \\
93.24628  & 18.04524 &  15.37 &  0.01 &  14.13 &  0.01 &  13.32 &  0.01 &  12.19 &  0.02 &  11.74 &  0.02 &  11.16 &  0.03 &  9.91 &  0.03 & II \\
93.15045  & 17.94815 &  14.27 &  0.01 &  13.18 &  0.01 &  12.53 &  0.01 &  11.51 &  0.02 &  11.09 &  0.02 &  10.94 &  0.04 &  9.96 &  0.08 & II \\
93.21329  & 18.01470 &  18.37 &  0.11 &  15.82 &  0.02 &  14.21 &  0.01 &  12.35 &  0.02 &  11.90 &  0.02 &  11.06 &  0.05 &  10.02 &  0.09 & II \\
93.23944  & 18.02592 &  17.47 &  0.06 &  16.02 &  0.03 &  14.76 &  0.02 &  13.61 &  0.04 &  12.34 &  0.03 &  11.67 &  0.04 &  10.12 &  0.05 & I \\
93.35432  & 17.88051 &  16.42 &  0.03 &  14.79 &  0.01 &  13.64 &  0.01 &  12.02 &  0.02 &  11.22 &  0.02 &  10.99 &  0.02 &  10.12 &  0.02 & II \\
93.16369  & 17.94083 &  16.80 &  0.03 &  14.95 &  0.01 &  14.13 &  0.01 &  12.84 &  0.04 &  12.35 &  0.03 &  11.59 &  0.11 &  10.12 &  0.15 & I \\
93.25438  & 17.96418 &  17.38 &  0.06 &  15.46 &  0.02 &  14.14 &  0.01 &  12.38 &  0.03 &  11.71 &  0.02 &  11.21 &  0.07 &  10.25 &  0.13 & II \\
93.36552  & 17.94602 &  16.45 &  0.02 &  14.93 &  0.01 &  13.94 &  0.01 &  12.67 &  0.03 &  12.14 &  0.02 &  11.51 &  0.04 &  10.26 &  0.06 & II \\
93.07682  & 17.98306 &  16.08 &  0.01 &  15.18 &  0.01 &  14.70 &  0.01 &  13.74 &  0.04 &  13.01 &  0.04 &  12.18 &  0.04 &  10.28 &  0.03 & I \\
93.11366  & 17.98991 &  17.55 &  0.04 &  16.29 &  0.03 &  15.21 &  0.01 &  13.81 &  0.04 &  13.36 &  0.04 &  12.05 &  0.05 &  10.46 &  0.05 & I \\
93.38181  & 17.93109 &  17.09 &  0.04 &  15.70 &  0.03 &  14.75 &  0.01 &  13.80 &  0.06 &  13.47 &  0.05 &  12.36 &  0.12 &  10.54 &  0.15 & I \\
93.16968  & 18.02927 &  16.04 &  0.01 &  15.04 &  0.01 &  14.56 &  0.01 &  13.80 &  0.05 &  13.63 &  0.05 &  12.15 &  0.09 &  10.56 &  0.10 & I \\
93.25941  & 18.00423 &  15.62 &  0.01 &  14.42 &  0.01 &  13.68 &  0.01 &  12.95 &  0.03 &  12.59 &  0.03 &  11.77 &  0.05 &  10.60 &  0.07 & II \\
93.22853  & 18.04282 &  16.54 &  0.02 &  15.42 &  0.02 &  14.60 &  0.01 &  13.35 &  0.03 &  12.70 &  0.03 &  11.83 &  0.04 &  10.63 &  0.06 & I \\
93.15380  & 17.94118 &  17.07 &  0.04 &  15.55 &  0.02 &  14.73 &  0.02 &  13.46 &  0.05 &  12.72 &  0.03 &  11.82 &  0.09 &  10.68 &  0.15 & I \\
93.26565  & 17.95645 &  16.58 &  0.03 &  15.08 &  0.01 &  14.42 &  0.01 &  13.60 &  0.04 &  13.43 &  0.05 &  12.29 &  0.08 &  10.74 &  0.08 & I \\
93.27636  & 17.98397 &  16.16 &  0.02 &  14.45 &  0.01 &  13.43 &  0.01 &  12.15 &  0.02 &  11.62 &  0.02 &  11.35 &  0.05 &  10.76 &  0.12 & II \\
93.12441  & 17.91705 &  18.71 &  0.10 &  15.58 &  0.01 &  13.93 &  0.01 &  12.36 &  0.02 &  11.99 &  0.02 &  11.12 &  0.02 &  10.79 &  0.03 & II \\
93.22486  & 18.01707 &  15.00 &  0.01 &  14.48 &  0.01 &  13.91 &  0.01 &  12.67 &  0.03 &  12.01 &  0.02 &  11.11 &  0.08 &  10.80 &  0.14 & II \\
93.24936  & 17.99069 &  16.45 &  0.02 &  15.15 &  0.01 &  14.57 &  0.01 &  13.46 &  0.04 &  13.03 &  0.04 &  11.83 &  0.14 &  10.80 &  0.17 & I \\
93.07597  & 17.87753 &  16.77 &  0.03 &  14.66 &  0.01 &  13.62 &  0.01 &  12.63 &  0.02 &  12.12 &  0.02 &  11.36 &  0.02 &  10.83 &  0.03 & II \\
93.27975  & 17.80454 &  16.30 &  0.02 &  14.91 &  0.01 &  13.88 &  0.01 &  12.73 &  0.03 &  12.32 &  0.03 &  11.46 &  0.02 &  10.84 &  0.03 & II \\
93.23428  & 18.04390 &  16.60 &  0.02 &  15.59 &  0.02 &  14.82 &  0.01 &  13.86 &  0.04 &  13.26 &  0.04 &  12.74 &  0.07 &  11.00 &  0.07 & I \\
93.11582  & 17.91519 &  18.00 &  0.05 &  16.11 &  0.02 &  15.06 &  0.01 &  13.40 &  0.03 &  13.24 &  0.04 &  11.82 &  0.03 &  11.08 &  0.04 & II \\
93.13181  & 17.93129 &  15.60 &  0.01 &  14.46 &  0.01 &  14.03 &  0.01 &  13.34 &  0.03 &  12.98 &  0.04 &  12.39 &  0.08 &  11.10 &  0.10 & II \\
93.12912  & 17.91860 &  18.47 &  0.09 &  16.15 &  0.02 &  14.73 &  0.01 &  13.62 &  0.04 &  12.91 &  0.03 &  11.80 &  0.03 &  11.15 &  0.03 & II \\
93.29075  & 18.04884 &  15.69 &  0.03 &  14.95 &  0.02 &  14.20 &  0.01 &  13.37 &  0.03 &  12.84 &  0.03 &  12.29 &  0.04 &  11.16 &  0.05 & II \\
93.20743  & 18.07821 &  15.33 &  0.02 &  14.17 &  0.01 &  13.41 &  0.01 &  12.60 &  0.02 &  12.24 &  0.02 &  11.79 &  0.03 &  11.18 &  0.04 & II \\
93.34914  & 17.88210 &  16.22 &  0.02 &  14.61 &  0.01 &  13.71 &  0.01 &  12.96 &  0.03 &  13.26 &  0.04 &  11.60 &  0.02 &  11.23 &  0.03 & II \\
93.21317  & 18.02341 &  16.64 &  0.06 &  15.59 &  0.04 &  14.65 &  0.02 &  13.47 &  0.04 &  13.04 &  0.04 &  12.36 &  0.09 &  11.27 &  0.16 & II \\
93.09181  & 17.84300 &  15.71 &  0.01 &  14.52 &  0.01 &  13.90 &  0.01 &  13.18 &  0.03 &  12.81 &  0.03 &  12.32 &  0.04 &  11.30 &  0.03 & II \\
93.03669  & 17.81515 &  16.58 &  0.03 &  15.38 &  0.02 &  14.48 &  0.02 &  13.47 &  0.04 &  12.54 &  0.03 &  12.46 &  0.04 &  11.31 &  0.03 & II \\
93.28318  & 18.05870 &  16.98 &  0.05 &  16.44 &  0.05 &  15.20 &  0.02 &  13.85 &  0.04 &  13.22 &  0.04 &  12.61 &  0.05 &  11.37 &  0.04 & II \\
93.18279  & 17.94522 &  16.60 &  0.02 &  15.00 &  0.01 &  14.12 &  0.01 &  12.94 &  0.03 &  12.60 &  0.03 &  12.09 &  0.05 &  11.39 &  0.10 & II \\
93.09937  & 17.95723 &  15.34 &  0.01 &  14.33 &  0.01 &  13.79 &  0.01 &  13.02 &  0.03 &  12.32 &  0.03 &  12.14 &  0.06 &  11.40 &  0.13 & II \\
93.26479  & 18.03973 &  17.04 &  0.03 &  16.04 &  0.03 &  15.18 &  0.02 &  14.04 &  0.05 &  13.53 &  0.05 &  12.68 &  0.10 &  11.40 &  0.17 & I \\
93.05625  & 17.98123 &  15.76 &  0.01 &  14.86 &  0.01 &  14.21 &  0.01 &  13.44 &  0.03 &  12.89 &  0.03 &  12.44 &  0.04 &  11.45 &  0.04 & II \\
93.08435  & 17.81195 &  16.03 &  0.01 &  14.71 &  0.01 &  14.12 &  0.01 &  13.48 &  0.04 &  12.86 &  0.03 &  12.30 &  0.04 &  11.46 &  0.03 & II \\
93.27053  & 17.85008 &  19.31 &  0.22 &  16.69 &  0.05 &  15.18 &  0.02 &  13.87 &  0.04 &  13.47 &  0.04 &  12.16 &  0.03 &  11.47 &  0.04 & I \\
93.12422  & 17.93572 &  15.90 &  0.01 &  14.74 &  0.01 &  14.23 &  0.01 &  13.21 &  0.03 &  13.03 &  0.04 &  12.28 &  0.04 &  11.49 &  0.07 & II \\
\enddata
\tablenotetext{a}{The entire table is available electronically, in the on-line version of this paper.}
\end{deluxetable}

%% file: tab3.tex
\begin{deluxetable}{lrccccrr} 
\tabletypesize{\footnotesize}
\tablecaption{Parameters of ionizing stars\label{ionizing_stars}}
\tablewidth{0pt}
\tablehead{
\colhead{Name} & 
\colhead{Name} & 
\colhead{RA (2000)} & 
\colhead{Dec (2000)} & 
\colhead{Spectral Type\tablenotemark{a}} & 
\colhead{Spectral Type} & 
\colhead{H\scriptsize{II} region age\tablenotemark{b}} &
\colhead{Reference}\\
& 
\colhead{Catalog}&
hh mm ss&
dd mm ss&
\colhead{Literature}& 
\colhead{Our data}& 
\colhead{[Yr]} 
}
\startdata
S254  & HD253247 & 06 12 22.1  & 18 00 57.7  & O9.0V & O9.6V  &  5.1$\times 10^6$ & \citet{mof79}  \\
S255  & HD253237 & 06 13 04.2  & 17 58 41.1  & B0.0V & B0.0V  &  1.5$\times 10^6$ & \citet{mof79}  \\
S256  &    -     & 06 12 36.5  & 17 56 54.3  & B2.5V & B0.9V  &  2$\times 10^5$   & \citet{rus07}  \\
S257  & ALS19    & 06 12 44.2  & 17 59 13.8  & B0.5V &   -    &  1.6$\times 10^6$ & \citet{mof79}  \\
S258  &    -     & 06 13 28.3  & 17 55 33.3  &   -   & B1.5V  &  1$\times 10^5$   &  - \\
\enddata         
\tablenotetext{a}{Spectral types from literature correspond to the reference listed in the last column of the table.}
\tablenotetext{b}{Ages were calculated using the spectral type given by our data. For S257 we used the spectral type from \citet{mof79}.}
\end{deluxetable}

%% file: tab4.tex
\begin{deluxetable}{lccrrrrlcrrrrrr} 
\tabletypesize{\scriptsize}
\tablecaption{Parameters of clusters\tablenotemark{a}\label{clusters}}
\tablewidth{0pt}
\tablehead{
\colhead{Name} & 
\colhead{RA (2000)} & 
\colhead{Dec (2000)} & 
\colhead{N$_{\textrm{\tiny{IR}}}$\tablenotemark{b}} & 
\colhead{N\tablenotemark{c}} & 
\colhead{I} & 
\colhead{II} & 
\colhead{I/II} & 
\colhead{I/N$_{\textrm{\tiny{IR}}}$} &
\colhead{$\sigma_{\textrm{\tiny{s}}}$\tablenotemark{d}} & 
\colhead{$\sigma_{\textrm{\tiny{smax}}}$\tablenotemark{e}} & 
\colhead{$\sigma_{\textrm{\tiny{g}}}$\tablenotemark{f}} & 
\colhead{$\sigma_{\textrm{\tiny{gmax}}}$\tablenotemark{g}} & 
\colhead{m$_{\textrm{\tiny{gas}}}$} &
\colhead{$\epsilon$} \\
 & 
\colhead{hh mm ss} & 
\colhead{dd mm ss} & 
 & & & & & &
\colhead{[pc$^{-2}$]} & 
\colhead{[pc$^{-2}$]} & 
\colhead{[cm$^{-2}$]} & 
\colhead{[cm$^{-2}$]} & 
\colhead{[M$_{\odot}$]} & 
\colhead{}
}
\startdata
S255-2 \& S255N    & 6 12 54 & 17 59 10 & 140 & 488 & 23  & 14  & 1.64\tablenotemark{h} &  0.16  & 18 & 156 & 2.9$\times 10^{16}$ & 7.0$\times 10^{16}$ & 1890 & 0.11\\
S256               & 6 12 36 & 17 56 48 & 115 & 276 & 14  & 44  & 0.32 &  0.12  & 16 & 256 & 1.6$\times 10^{16}$ & 3.6$\times 10^{16}$ & 2005 & 0.06\\
S258               & 6 13 29 & 17 55 44 &  49 & 186 &  7  &  9  & 0.78 &  0.14  & 17 & 298 & 1.4$\times 10^{16}$ & 2.3$\times 10^{16}$ &  815 & 0.10\\
G192.54-0.15       & 6 12 24 & 17 59 27 &   9 &  88 &  2  &  0  &  -   &  0.22  & 17 & 166 & 2.0$\times 10^{15}$ & 4.1$\times 10^{15}$ &   40 & 0.54\\
G192.75-0.00       & 6 13 24 & 17 52 51 &   7 &  60 &  0  &  7  & 0.00 &  0.00  & 12 &  76 & 8.0$\times 10^{15}$ & 9.2$\times 10^{15}$ &  100 & 0.23\\
G192.75-0.08       & 6 13 05 & 17 50 30 &  25 &  50 &  9  & 15  & 0.60 &  0.36  & 11 & 238 & 1.2$\times 10^{16}$ & 1.9$\times 10^{16}$ &  490 & 0.05\\
G192.63-0.00       & 6 13 08 & 17 58 42 &  22 &  44 &  7  &  5  & 1.40\tablenotemark{h} &  0.32  & 10 &  39 & 1.4$\times 10^{16}$ & 2.0$\times 10^{16}$ &  560 & 0.04\\
G192.69-0.25       & 6 12 19 & 17 48 28 &   7 &  14 &  3  &  4  & 0.75 &  0.43  &  9 &  27 & 1.0$\times 10^{16}$ & 1.4$\times 10^{16}$ &  125 & 0.05\\
G192.65-0.08       & 6 12 53 & 17 55 30 &  13 &  26 &  2  &  6  & 0.33 &  0.15  &  9 &  37 & 1.8$\times 10^{16}$ & 3.0$\times 10^{16}$ &  340 & 0.04\\
G192.55-0.01       & 6 12 56 & 18 02 44 &   9 &  18 &  2  &  5  & 0.40 &  0.22  &  9 &  30 & 2.3$\times 10^{15}$ & 4.8$\times 10^{15}$ &   20 & 0.33\\
\enddata
\tablenotetext{a}{We assumed a distance of 2.4 kpc from the Sun in the estimation of physical parameters.}
\tablenotetext{b}{Number of stars with IR-excess, including Class I and Class II.}
\tablenotetext{c}{Corrected number of cluster members, see \S~\ref{section_efficiency}.}
\tablenotetext{d}{Mean surface density of stars.}
\tablenotetext{e}{Peak surface density of stars.}
\tablenotetext{f}{Mean $^{13}\mathrm{CO}$ surface density.}
\tablenotetext{g}{Peak $^{13}\mathrm{CO}$ surface density.}
\tablenotetext{h}{Likely to have missed some Class II (see end of \S~\ref{7band}).}
\end{deluxetable}

%% file: ms.bbl
\begin{thebibliography}{}
\bibitem[Allen et al.(2004)]{all04} Allen, L. E., Nuria, C., et al., 2004, ApJS, 154, 363
\bibitem[Allen et al.(2007)]{all07} Allen, L. E., Megeath, T., et al., 2007, PPV conference, 361A 
\bibitem[Andr\'e et al.(2000)]{and00} Andr\'e, P., Ward-Thompson, D., Barsony, M., 2000, ``Protostars and Planets IV'', p59
\bibitem[Beichman et al.(1979)]{bei79} Beichman, C. A., Becklin, E. E., \& Wynn-Williams, C., 1979, ApJ, 232, L47
\bibitem[Bessel \& Brett(1988)]{bes88} Bessel, M., \& Brett, J., 1988, PASP, 100, 1134
\bibitem[Bica et al.(2003)]{bic03} Bica E., Dutra C., M., Soares, J., Barbuy, B., 2003, A\&A, 404, 223B
\bibitem[Brand \& Blitz(1993)]{bra93} Brand, J., \& Blitz, L., 1993, A\&A, 275, 67
\bibitem[Carpenter et al.(1995)]{car95} Carpenter, J., M., Snell, R., L., Schloerb F., P., 1995, ApJ, 445, 246
\bibitem[Chavarr\'{\i}a-K et al.(1987)]{cha87} Chavarr\'{\i}a-K, C., de Lara, E., Hasse, I., 1987, A\&A 171, 216
\bibitem[Chapman et al.(2007)]{cha07} Chapman, N., L., Lai, S., Mundy, L., G., et al., 2007, ApJ 667, 288
\bibitem[Cushing et al.(2004)]{cus04} Cushing, M., C., Vacca, W., D., \& Rayner, J., T., 2004, PASP, 116, 362C
\bibitem[Cyganowski et al.(2007)]{cyg07} Cyganowski, C., J., Brogan, C., L., Hunter, T., R., 2007, AJ, 134, 346
\bibitem[Dutra \& Bica(2001)]{dut01} Dutra C., M., \& Bica E., 2001, A\&A, 376, 434D
\bibitem[Eisenhardt et al.(2004)]{eis04} Eisenhardt, P., R., Stern, D., et al., 2004, ApJS, 154, 48
\bibitem[Evans et al.(1977)]{eva77} Evans, N., Blair, G. N., \& Beckwith, S. 1977, ApJ, 217, 448
\bibitem[Elmegreen et al.(2000)]{elm00} Elmegreen, B. G., Efremov, Y., et al., 2000, in Protostars and Planets IV, ed. V. Mannings, A. P. Boss, \& S. S. Russell (Tucson: Univ. of Arizona Press), 179
\bibitem[Erickson et al.(1999)]{eri99} Erickson, N. R., Grosslein, R. M., Erickson, R. B., \& Weinreb, S. 1999, IEEE Trans. Microwave Theory Tech., 47, 2212
\bibitem[Flaherty et al.(2007)]{fla07} Flaherty, K., M., Pipher, J., M., et al., 2007, ApJ, 663, 1069F
\bibitem[Frerking et al.(1982)]{fre82} Frerking, M., A., Langer, W., D., \& Wilson, R., W., 1982, ApJ, 262, 590
\bibitem[Greene et al.(1994)]{gre94} Greene, T., P., et al. 1994 ApJ, 434, 614
\bibitem[Gutermuth et al.(2007)]{gut07} Gutermuth, R., A., Myers, P., C., et al., 2007, arXiv, 0710, 1860G
\bibitem[Harvey et al.(2007a)]{har07} Harvey, P., M., et al., 2007a, ApJ, 663, 1149
\bibitem[Harvey et al.(2007b)]{har07b} Harvey, P., M., Chapman, N., Lai, S., et al., 2007b, ApJ, 644, 307
\bibitem[Heyer et al.(1989)]{hey89} Heyer, M. H., Snell, R. L., Morgan, J., Schloerb, F. P. 1989, ApJ, 346, 220
\bibitem[Heyer et al.(1998)]{hey98} Heyer, M. H., Brunt, C. M., Snell, R. L., Howe, J., Schloerb, F. P., \& Carpenter, J. C., 1998, ApJS, 115, 241
\bibitem[Hora et al.(2004)]{hor04} Hora, J., L., et al., 2004, SPIE, 5487, 77
\bibitem[Hunter \& Massey(1990)]{hun90} Hunter, D., A., Massey, P., 1990, AJ, 99, 84
\bibitem[Indebetouw et al.(2005)]{ind05} Indebetouw, R., Mathis, J., S., et al., 2005, ApJ, 619, 931
\bibitem[Israel et al.(2003)]{isr03} Israel, F., P., de Graauw, Th., et al., 2003, A\&A, 401, 99
\bibitem[Jaffe et al.(1984)]{jaf84} Jaffe, D.,T., et al., 1984, ApJ, 284, 637
\bibitem[J\o rgensen et al.(2006)]{jor06} J\o rgensen, J., K., Harvey, P., M., et al., 2006, ApJ, 645, 1246.
\bibitem[Kumar et al.(2007)]{kum07} Kumar, M., N., S., Davis, C., J., et al. 2007, MNRAS, 374, 54
\bibitem[Kutner \& Ulich(1981)]{kut81} Kutner, M. L., \& Ulich, B. L., 1981, ApJ, 250, 341
\bibitem[Lada(1987)]{lad87} Lada, C., J., 1987, IAUS, 115, 1L  	
\bibitem[Lada et al.(2006)]{lad06} Lada, C., et al., 2006, AJ, 131, 1574
\bibitem[Lo \& Burke(1973)]{lo73} Lo, K Y., \& Burke, B. F. 1973, A\&A, 26, 487
\bibitem[Minier et al.(2005)]{min05} Minier V., et al., 2005, A\&A, 429, 945
\bibitem[Miralles et al.(1997)]{mir97} Miralles M. P., Salas L., Cruz-Gonz\'alez I., Kurtz S., 1997, ApJ, 488, 749
\bibitem[Mizuno(1982)]{miz82} Mizuno, S., 1982, Ap\&SS, 87, 121
\bibitem[Moffat et al.(1979)]{mof79} Moffat, A., FitzGerald, M., P., \& Jackson, P., D., 1979, A\&AS, 38, 197
\bibitem[Muench et al.(2000)]{mue00} Muench, A., A., Lada, E., A., \& Lada, C., J., 2000, ApJ, 533, 358
\bibitem[Padgett et al.(2008)]{pad08} Padgett, D., L., Rebull, L., M.,  Stapelfeldt, K., R., et al., 2008, ApJ, 672, 1013
\bibitem[Patten et al.(2006)]{pat06} Patten, B., Stauffer, J., R., et al., 2006, ApJ, 651, 502
\bibitem[P\'equignot et al.(1991)]{peq91} P\'equignot, D., Petitjean, P., \& Boisson, C., 1991, A\&A, 251, 680
\bibitem[Pismis \& Hasse(1976)]{pis76} Pismis, P., \& Hasse, I., 1976, ApSS, 45, 79
\bibitem[Pismis \& Hasse(1982)]{pis82} Pismis, P., \& Hasse, I., 1982, RMxAA, 5, 161
\bibitem[Porras et al.(2007)]{por07} Porras, A., J\o rgensen , J., Allen, L., et al., 2007, ApJ, 656, 493
\bibitem[Rayner et al.(2003)]{ray03} Rayner, J., T., 2003, PASP, 115, 362R
\bibitem[Reed(2005)]{ree05} Reed, C., 2005 yCat., 5125
\bibitem[Reid(2008)]{rei08} Reid, M., 2008, Proceedings IAU Symposium 248
\bibitem[Robitaille et al.(2006)]{rob06} Robitaille, T., et al., 2006 ApJS, 167, 256
\bibitem[Russeil et al.(2007)]{rus07} Russeil, D., Adami, C., \& Georgelin, Y., M., 2007, A\&A, 470, 161
\bibitem[Schuster et al.(2006)]{sch06} Schuster, M., T., Marengo M., Patten, B., M., 2006, SPIE, 6270, 65
\bibitem[Sharpless(1959)]{sha59} Sharpless, S. 1959, ApJ Suppl., 4, 257
\bibitem[Stahler \& Palla(2004)]{sta04} Stahler Steven W., Palla Francesco, ``The Formation Of Stars'', 2004, Wiley VCH
\bibitem[Turner(1971)]{tur71} Turner, B. E. 1971, ApJ Letters, 8, 73
\bibitem[Vacca et al.(2003)]{vac03} Vacca, W., D., Cushing, M., C., \& Rayner, J., T., 2003, PASP, 115, 389V
\bibitem[Wright et al.(1981)]{wri81} Wright, E. L., Harper, D. A., Loewenstein, R. F., \& Moseley, H., 1981, ApJ, 246, 426
\end{thebibliography}
